%% file: main.tex
\documentclass[sigconf]{acmart}
\usepackage{enumitem}
\AtBeginDocument{%
  }

\copyrightyear{2026}
\acmYear{2026}
\setcopyright{cc}
\setcctype{by}
\acmConference[CHI '26]{Proceedings of the 2026 CHI Conference on Human Factors in Computing Systems}{April 13--17, 2026}{Barcelona, Spain}
\acmBooktitle{Proceedings of the 2026 CHI Conference on Human Factors in Computing Systems (CHI '26), April 13--17, 2026, Barcelona, Spain}
\acmPrice{}
\acmDOI{10.1145/3772318.3791552}
\acmISBN{979-8-4007-2278-3/2026/04}

\begin{document}

\title{Compliant But Unsatisfactory: The Gap Between Auditing Standards and Practices for Probabilistic
Genotyping Software}

\author{Angela Jin}
\affiliation{%
  \institution{University of California, Berkeley}
  \city{Berkeley}
  \state{California}
  \country{USA}
}
\email{angelacjin@berkeley.edu}

\author{Alexander Asemota}
\affiliation{%
  \institution{University of California, Berkeley}
  \city{Berkeley}
  \state{California}
  \country{USA}
}
\email{alexander.asemota@berkeley.edu}

\author{Dan E. Krane}
\affiliation{%
  \institution{Wright State University}
  \city{Dayton}
  \state{Ohio}
  \country{USA}
}
\email{dan.krane@wright.edu}

\author{Nathaniel Adams}
\affiliation{%
  \institution{Forensic Bioinformatic Services, Inc.}
  \city{Fairborn}
  \state{Ohio}
  \country{USA}
}
\email{adams@bioforensics.com}

\author{Rediet Abebe}
\affiliation{%
  \institution{ELLIS Institute; Max Planck Institute for Intelligent Systems; and T{\"u}bingen AI Center}
  \city{T{\"u}bingen}
  \country{Germany}
}
\email{aim-admin@tue.ellis.eu}

\renewcommand{\shortauthors}{Jin, Asemota, Krane, Adams, and Abebe}

\begin{abstract}
AI governance efforts increasingly rely on audit standards: agreed-upon practices for conducting audits. However, poorly designed standards can hide and lend credibility to inadequate systems. We explore how an audit standard’s design influences its effectiveness through a case study of ASB 018, a standard for auditing probabilistic genotyping software---software that the U.S. criminal legal system increasingly uses to analyze DNA samples. Through qualitative analysis of ASB 018 and five audit reports, we identify numerous gaps between the standard's desired outcomes and the auditing practices it enables. For instance, ASB 018 envisions that compliant audits establish restrictions on software use based on observed failures. However, audits can comply without establishing such boundaries. We connect these gaps to the design of the standard’s requirements such as vague language and undefined terms. We conclude with recommendations for designing audit standards and evaluating their effectiveness.
\end{abstract}




% COMMENTED OUT FOR SUBMISSION
% \received{20 February 2025}

\maketitle

\input{1-introduction}

\input{2-background}

\input{3-related-work}

\input{4-methods}

\input{5-results}
\input{6-discussion}
\input{7-conclusion}

\begin{acks}
Thank you to Marc Canellas, Finale Doshi-Velez, Krzysztof Gajos, Jeanna Matthews, Andrea Roth, Rebecca Wexler, and the Harvard Human-Computer Interaction group for conversations that motivated and shaped this work. Thank you to Jennifer Friedman, Inioluwa Deborah Raji, Niloufar Salehi, and our anonymous reviewers for their comments and feedback on earlier versions of this paper. This material is based upon work supported by the National Science Foundation Graduate Research Fellowship Program under Grant No. DGE 2146752 and by the National Science Foundation under Grant No. 2243822. This research was also supported by the Andrew Carnegie Fellows Program, and the Hector Foundation. Any opinions, findings, and conclusions or recommendations expressed in this material are those of the authors and do not necessarily reflect the views of the National Science Foundation or other funding agencies. 
\end{acks}

\bibliographystyle{ACM-Reference-Format}
\bibliography{sample-base}

\appendix
\input{8-appendix}

\end{document}

%% file: 1-introduction.tex
\section{Introduction}
\begin{figure*}[h!]
    \centering
    \includegraphics[width=\textwidth]{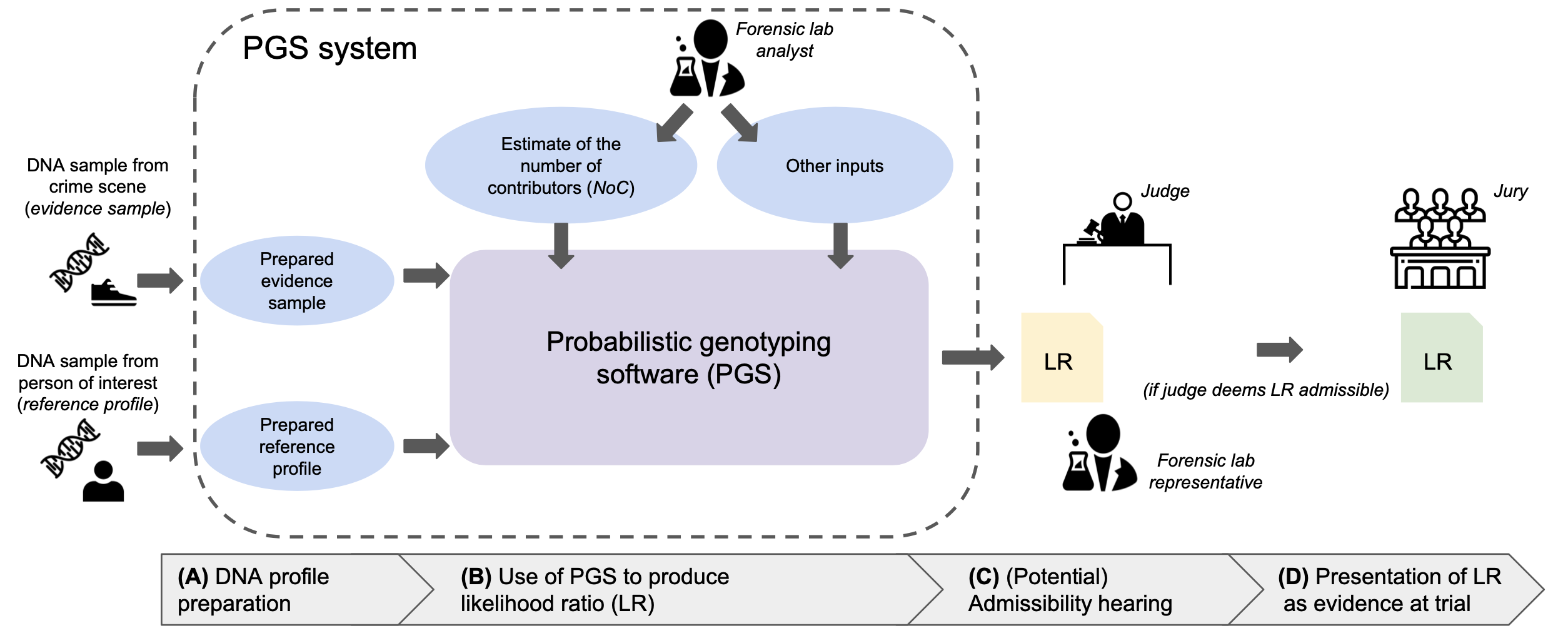}
    \caption{A simplified depiction of forensic lab use of probabilistic genotyping software to produce evidence in a criminal case. The goal of forensic DNA testing is to determine whether a person of interest (e.g., the defendant) could be the source of some or all of the DNA recovered from a crime scene. To this end, \textbf{(A)} a lab analyst first prepares two physical DNA samples: an evidence sample (e.g., a swabbing of a shoe collected as evidence) and a reference profile. \textbf{(B)} The analyst then runs the PGS with these DNA samples as inputs, along with other inputs chosen by the analyst. The software outputs a \textit{likelihood ratio} (LR)---a statistic that describes similarities observed between the evidence sample and reference profile. \textbf{(C)} Before the LR is presented as evidence at trial, a court may conduct an admissibility hearing to determine whether the evidence is sufficiently reliable to be presented in front of a jury. \textbf{(D)} If the judge deems the evidence admissible, the LR will be presented as evidence at trial.}.
    \label{fig:pgs-flowchart}
    \Description{Flow-chart depicting stages of the forensic DNA testing process and the introduction of the probabilistic genotyping software (PGS) likelihood ratio output as evidence in a criminal case. The flowchart has four stages: (A) DNA profile preparation, (B) Use of PGS to produce likelihood ratio, (C) A potential admissibility hearing, and (D) Presentation of LR as evidence at trial. The figure depicts four inputs to the probabilistic genotyping software: the crime scene DNA sample (i.e., evidence sample), the person of interest's DNA sample (i.e., reference profile), the forensic lab analyst's estimate of the number of contributors, and other analyst inputs. The software outputs a likelihood ratio (LR), which the figure then depicts as being presented in front of a judge. If the judge deems the LR admissible, the LR is shown as trial evidence in front of a jury. There is a box depicting the PGS. There is also a larger box encompassing both the PGS and lab analyst with a dotted line depicting the broader PGS system that takes physical DNA samples as input and outputs a LR.}
\end{figure*}

Over the past decade, researchers and practitioners have developed an ecosystem of tools---including guidelines, frameworks, and software toolkits---to promote fairness, transparency, and accountability throughout the AI lifecycle~\cite{madaio2020co, holstein2019improving, lee2021landscape, raji2020closing, mitchell2019model, abebe2022adversarial}. Amidst growing calls for algorithm audits, recent work highlights that such tools can be crucial for ensuring these evaluations result in meaningful accountability~\cite{ojewale2025towards, abebe2022adversarial, lam2024framework, raji2022outsider, jin2024beyond}. However, tools for responsible AI may fail to achieve their intended outcomes when deployed in complex organizational settings and practices~\cite{wong2023seeing, berman2024scoping, deng2022exploring, rakova2021responsible, madaio2020co, schor2024mind}.

In response, a burgeoning body of work investigates how to design audit tooling that effectively helps audits create accountability~\cite{ojewale2025towards, madaio2020co, raji2020closing, holstein2019improving, lam2024framework, raji2022outsider, abebe2022adversarial}. As researchers and policymakers increasingly call for audit standards (i.e., agreed-upon best practices)~\cite{goodman2022algorithmic, costanza2022audits, lam2024framework, raji2022outsider, groves2024auditing, aistandardshub}, we ask: \textit{How can the design of an audit standard undermine its effectiveness?} We specifically examine how a standard’s formal requirements may enable audits that comply with the letter of the standard while failing to meet the goals envisioned by the standard's developers. We explore this question through the lens of \textbf{ASB 018\footnote{We study the first edition of ASB 018, which was developed and published in 2020 by the American Academy of Forensic Sciences Standards Board (ASB)~\cite{asb018}.}}, a standard for auditing \textbf{probabilistic genotyping software (PGS)} used by U.S. forensic laboratories to analyze DNA samples in criminal cases. ASB 018 provides a critical case study not only because it mirrors audit processes in other domains like hiring~\cite{raji2022outsider, lam2024framework, groves2024auditing, gerchick2025auditing}, healthcare~\cite{obermeyer2019dissecting, wu2021medical}, and social services~\cite{chouldechova2018case} but also because forensic labs already cite ASB 018 compliance to convince judges that PGS outputs are sufficiently reliable to be presented as evidence in criminal trials~\cite{2023anderson}.

We investigate ASB 018's design and effectiveness through two research questions: \textit{(RQ1) What gaps exist between the practices ASB 018 envisions and the practices it enables?\footnote{We study forensic labs' audit practices by examining their representation in publicly available audit reports; see Section~\ref{sec:methods}.}}, and \textit{(RQ2) How does the design of ASB 018 facilitate these gaps?} In this work, we analyzed five publicly available PGS audit reports spanning various lab sizes, software versions, and years. We first assessed each audit report for compliance with ASB 018’s formal requirements. After determining that all five audits could be interpreted as compliant, we qualitatively coded ASB 018 and its accompanying factsheet to identify its desired outcomes across five audit stages~\cite{ojewale2025towards}: Audit Scope, Standards Identification, Performance Analysis, Post-Audit Judgment, and Audit Communication. We then qualitatively coded each audit report to identify practices that misalign with the standard’s goals (RQ1). Finally, we conducted a second round of coding on ASB 018, this time focusing on the language and organization of its requirements to identify aspects of its design that allow unsatisfactory audits to remain technically compliant (RQ2).

Our findings reveal that the analyzed audits are ASB 018 \textit{compliant, but unsatisfactory}: \textbf{audit practices consistently fall short of ASB 018’s envisioned goals across all five audit stages, yet still satisfy the standard's formal requirements}. For instance, while ASB 018 envisions testing the broader sociotechnical system, the audits primarily measure narrow software performance. Similarly, although the standard intends for audits to establish boundaries on software use based on failures observed during performance analysis, the audits avoid framing errors as failures and establish no such boundaries. Furthermore, despite the standard's emphasis on documentation that supports third-party evaluation, the reports omit crucial details needed to verify audit claims. \textbf{ASB 018 enables these gaps through design choices such as undefined concepts, requirements language that treats audit components as separate considerations to be addressed in isolation, and vague language calling for labs to \textit{``address''} and \textit{``consider''} that allow compliance with minimal to no testing.}

To conclude, we discuss how our study demonstrates that clear articulation of a tool’s goals---even from an external perspective---is essential for evaluating and improving audit infrastructure. Towards designing audit tools that more effectively deliver accountability, we caution that designing for compatibility with users' current perspectives and practices can undermine the tool's effectiveness. To help navigate this tension, we provide recommendations for those designing standards to (1) evaluate standards against fine-grained articulations of desired outcomes to identify where increased specificity is needed, using frameworks such as that by \citet{ojewale2025towards} to articulate intended outcomes for audit stages beyond evaluation; (2) design requirements that clearly specify actor responsibilities, minimum activities, and intermediary steps essential to achieving desired outcomes; and (3) co-design standards with stakeholders beyond intended users to encourage more robust definitions of desired outcomes that counterbalance users' desires for flexibility.

%% file: 2-background.tex
\section{Background}
\label{sec:background}

\subsection{Probabilistic Genotyping Software Systems}
\label{sec:background-pgs}
Forensic DNA testing aims to determine whether a person of interest (e.g., the defendant) could be the source of some or all of the DNA recovered from a crime scene~\cite{butler2024dna}. Since 2009, the U.S. criminal legal system has increasingly relied on \textbf{probabilistic genotyping software (PGS)} to analyze complex DNA evidence~\cite{moss2015admissibility, foley}.\footnote{\citet{moss2015admissibility} describes, ``In 2009, TrueAllele made its first appearance in a United States courtroom in the murder case of Commonwealth v. Foley.''}. PGS are now widely used in criminal cases around the country. For example, in March of 2025, the creators of STRmix---one of the two leading PGS alongside TrueAllele---announced that STRmix is being used by 91 forensic labs across the U.S., by 29 more labs internationally, and in over 690,000 criminal cases worldwide~\cite{strmix2025use}

Figure~\ref{fig:pgs-flowchart} depicts how forensic labs integrate PGS in casework. A laboratory analyst first prepares two DNA samples for input into the software: an \emph{evidentiary sample} taken from the crime scene and a \emph{reference profile} from a person of interest. The analyst then provides these samples as inputs to the software alongside subjective parameters, most notably the analyst's best estimate for the ``apparent number of contributors.'' The PGS uses these inputs to produce a \textbf{likelihood ratio (LR)}. This final output is significantly impacted by the analyst's decision-making, such as their estimate for the number of contributors, as well as upstream laboratory equipment~\cite{butler2024dna, paoletti2005empirical,coble2015uncertainty}.\footnote{For more background on likelihood ratios and their use in forensic DNA statistics, see Section 2.5, ``Likelihood Ratios: Introduction to Theory and Application,'' in \citet{butler2024dna}}.

The LR compares the probability of observing the evidentiary sample under two competing hypotheses: one where the person of interest is a contributor and one where they are not.\footnote{For example, an LR may compare a hypothesis that the person of interest and two unknown individuals contributed to the DNA mixture, against a hypothesis that three unknown individuals who are not the person of interest contributed to the DNA mixture.} LR values greater than 1 are interpreted as providing support for \textbf{inclusion}, while values less than 1 support \textbf{exclusion}. An $LR = 1$ is considered ``uninformative.'' We define a false negative result as an exclusionary LR ($LR < 1$) for a true contributor (i.e., a person of interest whose DNA \textit{is} in the evidentiary sample) and a false positive as an inclusionary LR ($LR > 1$) for a non-contributor (i.e., a person of interest whose DNA is \textit{not} in the evidentiary sample).

\subsection{Validation of PGS Systems}
\label{sec:background-pgs-validation}
In forensics, validation is the empirical testing of a method prior to its deployment in casework. For PGS, this process consists of two parts: \emph{developmental validation} (conducted by software developers) and \emph{internal validation} (conducted by forensic laboratories seeking to use the software in casework)~\cite{butler2024dna, coble2019probabilistic}. In this work, we focus on internal validation studies, which we conceptualize as \textit{audits}\footnote{In this paper, we use \citet{birhane2024ai}'s definition of an audit as ``any independent assessment of an identified audit target via an evaluation of articulated expectations with the implicit or explicit objective of accountability.'' For a detailed explanation of the terms used in this definition, see Section II.A in \citet{birhane2024ai}.}.

\subsubsection{Purpose of internal validation studies} 
A forensic lab must complete an internal validation study before using a PGS in casework. A goal of internal validation is to establish the operational range---the range of DNA profile characteristics for which the lab has demonstrated acceptable performance~\cite{butler2024dna}. Dozens of profile characteristics are known to impact the reliability of PGS. Examples of those known to have the greatest impact include the number of contributors to the sample (NoC), the ratio of DNA amounts of contributors to the sample (if the sample is a DNA mixture, i.e., has more than one contributor), and the total amount of DNA in the sample~\cite{butler2024dna}.\footnote{For a list of notable profile characteristics, see Table 4.1 in \citet{butler2024dna}.}

Internal validation studies and the operational range established using study results can play a critical role in judges' decisions about whether or not the PGS outputs are sufficiently reliable to be presented as evidence at trial~\cite{roth2016machine}. This has been the case in several federal courts. For instance, the courts in U.S. v. Ortiz~\cite{ortiz} and U.S. v. Johnston~\cite{johnston} excluded PGS outputs due to sample characteristics in these cases exceeding the operational bounds established by the forensic laboratories' internal validation studies. Likewise, the court in U.S. v. Lewis~\cite{2020lewis} admitted some evidence and excluded some evidence after similar considerations. An internal validation study's compliance with an established standard can provide additional assurance to a judge that the forensic lab performed a high quality internal validation study and that the operational range is reliable.

\subsection{Standards and Guidelines for PGS Validation}
\label{sec:background-pgs-guidance}
In 2009, the National Academy of Sciences issued a report identifying a wide range of shortcomings in U.S. forensic science research and practice~\cite{national2009strengthening}. Recommendations six to eight of this report emphasize the need to establish forensic science standards, to accredit laboratories adhering to those standards, and to establish quality assurance and quality control practices for individual practitioners in order to promote rigor and reliability in forensic science methods. 

Following the NAS report, standards have been developed for a range of forensic disciplines. However, recent work raises the concern that these efforts have not yielded meaningful improvements and instead threaten to maintain the status quo~\cite{sinha2022radically, morrison2020vacuous}. We build on this work by exploring how standards for internal validation of probabilistic genotyping software may contain similar deficiencies. While several national and international groups have issued guidance for PGS validation~\cite{swgdam2015guidelines, coble2016dna, ukfsr2024guidelines, enfsi2023guidelines, fbi2025qas}, unlike ASB 018, these standards and guidelines are not approved by or registered with the American National Standards Institute (ANSI).

\subsubsection{The AAFS Standards Board (ASB)} 
\label{sec:background-asb}
In 2015, the American Academy of Forensic Sciences established the \textbf{AAFS Standards Board (ASB)}, a standards development organization tasked with developing standards for forensics practice in the U.S.\footnote{For more background on the creation of the ASB, see Appendix~\ref{appendix-a}.} Importantly, as opposed to the guidelines referenced in the previous paragraph, ASB standards are approved by and registered with the ANSI. This means that the ASB develops standards following ANSI requirements for ``openness, balance, lack of dominance, due process, and consensus''~\cite{ansi}. These requirements seek to ensure that the standards development process is ``equitable, accessible, and responsive to the requirements of various stakeholders,'' and that standards developed by the ASB allow organizations to ``maintain autonomy from vested interest groups''~\cite{asb2025about}.

\subsubsection{ASB Standard 018} 
\label{sec:background-asb018}
In 2020, the ASB published the first edition of \textbf{ASB 018, the Standard for Validation of Probabilistic Genotyping Systems}\footnote{ASB 018 is publicly available for download at: \url{https://www.aafs.org/sites/default/files/media/documents/018_Std_e1.pdf}}, which outlines a set of requirements for developmental and internal validations of PGS~\cite{asb018}. 

ASB 018 can influence internal validation study practices in several ways. On one hand, it can provide guidance to labs conducting internal validation studies. On the other hand, it can shape internal validation studies through external scrutiny that pressures labs to ensure internal validation compliance with the standard. The latter currently happens primarily through admissibility hearings in criminal cases, where a judge may interpret a lab's claim of compliance with ASB 018 as providing assurance of validation study quality.\footnote{When PGS is introduced in a criminal case in the U.S., the court may hold an admissibility hearing to determine whether the evidence is sufficiently reliable to be presented in front of the jury. All federal courts and some state courts use the Daubert Standard to conduct this assessment, and one factor considered under the Daubert Standard is \textit{``the existence and maintenance of standards controlling its operation.''}~\cite{wexlaw_daubert} The U.S. v. Anderson court conducted a Daubert hearing for one PGS software, and concluded: ``\textit{the Government has established that [the PGS tool] complies with the relevant standards issued by [SWGDAM] and ANSI/ASB. Accordingly, this factor weighs in favor of admissibility}''~\cite{2023anderson}.} 

In this paper, we focus ASB 018's effectiveness as a tool for assessing and providing assurance of internal validation study quality. Our investigation of ASB 018's effectiveness in providing appropriate assurance about compliant studies builds on critical concerns raised by recent work highlighting that judges frequently hold uncritical perceptions of forensic software~\cite{jin2024beyond} and work raising concerns that standards create ``a danger [...] that a court may not look further than the fact that a standard exists, and be misled into believing that conformity to a vacuous standard is indicative of scientific validity, even though it is not''~\cite[p.207]{morrison2020vacuous}.

%% file: 3-related-work.tex
\section{Related Work}
\label{sec:relatedwork}

\subsection{Audits and Accountability}
\label{sec:relatedwork-audits}
Researchers, law and policymakers, and civil society groups increasingly call for audits as mechanisms for identifying and mitigating risks of algorithmic systems and holding developers and users accountable for the algorithms they develop and use~\cite{raji2022outsider, wyden2023algorithmic}. \citet{birhane2024ai} define an audit as \textit{``any independent assessment of an identified audit target via an evaluation of articulated expectations with the implicit or explicit objective of accountability.''} Audits are not restricted to a specific method or type of audit target~\cite{ojewale2025towards, birhane2024ai, goodman2022algorithmic}. An audit may focus narrowly on a specific model or technical system~\cite{buolamwini2018gender}, or more broadly on the sociotechnical system encompassing technical components and human users~\cite{lam2023sociotechnical}. Audits may assess computational notions of fairness, but may also focus on other criteria such as functionality and legality~\cite{radiya2023sociotechnical, ojewale2025towards, abebe2022adversarial}. Additionally, audits may be conducted by a variety of actors, including an entity within the organization that built the examined algorithmic system, or those outside the organization. Most importantly, an audit does not end with a thorough performance analysis. Instead, evaluation of the audit target is situated within a broader accountability process in which the audited entity or organization faces consequences or is otherwise responsible for acting on the outcomes of the evaluation~\cite{ojewale2025towards}.

However, audits do not always successfully create accountability~\cite{latonero2021human, selbst2021institutional, goodman2022algorithmic, raji2022outsider}. When an audit is poorly designed or executed, a claim that a system has been audited can, at best, be meaningless, and at worst, lead to ``audit washing'' where the audit provides false assurance by lending credibility to a dysfunctional or otherwise inadequate system~\cite{goodman2022algorithmic}. For example, recent studies investigating audits of AI hiring systems performed in accordance with New York City's Local Law 144 (LL144) discuss how audits might be able to comply with the law's audit requirements despite being incomplete assessments of algorithmic bias~\cite{wright2024null, gerchick2025auditing}. Collectively, this body of work calls for careful attention to how audits are designed and conducted, and highlights how inadequate audits can hide and lend credibility to dysfunctional or inadequate AI systems. We build on this work by exploring how the design of an auditing standard can contribute to audit washing by lending credibility to audits that can be interpreted to comply with the standard despite failing to produce the outcomes envisioned by the standard's creators.

\subsection{Designing Effective Tools for Responsible AI}
A growing body of work in human-computer interaction and computer-supported cooperative work seeks to design tools such as software, frameworks, and checklists to help various stakeholders promote fairness, transparency, and accountability throughout the development, deployment, and use of AI systems~\cite{deng2022exploring, lee2021landscape,madaio2020co,raji2020closing, gebru2021datasheets, mitchell2019model}. Recent work examines these tools in the context of AI auditing, highlighting tools that support various stages of the auditing process such as audit standards that outline best practices, and repositories that facilitate broader communication of audit results~\cite{ojewale2025towards}. Several groups have explicitly highlighted the importance of audit standards as tools for improving audit quality and consistency, thereby minimizing the risk of audit washing~\cite{costanza2022audits, raji2022outsider, goodman2022algorithmic, ojewale2025towards, lam2024framework}.

However, numerous studies have shown that the existence of an RAI tool, alone, does not guarantee its effectiveness. A tool must be adopted and used, often within organizational structures and cultures that can hinder a tool's effectiveness in practice~\cite{wong2023seeing, madaio2020co, rakova2021responsible, metcalf2021algorithmic}. Collectively, this body of work emphasizes the importance of examining how a tool is (or might be) used in practice and designing tools with real-world organizational cultures, settings, and incentives in mind. In parallel and in response to these studies, recent work has called for the human-computer interaction community to expand its work to evaluating the effectiveness, and not just the usability, of these tools~\cite{berman2024scoping}.

\subsubsection{Designing Effective Audit Standards}
 Similarly, the existence of an audit standard, or an audit's compliance with a standard, does not necessarily ensure accountability~\cite{ojewale2025towards}. In fact, poorly designed standards can lead to ineffective audits and can lend further credibility to inadequate audits. Standards defined by the audit target~\cite{goodman2022algorithmic, hoofnagle2016assessing}, and standards with overly broad or vague language~\cite{goodman2022algorithmic, costanza2022audits, raji2022outsider, wright2024null, morrison2020vacuous}, can fail to ensure adequate audits and can add further credibility to an inadequate system. For example, \citet{wright2024null} examine audit requirements defined in New York City's Local Law 144 and discuss how the language grants companies substantial discretion over whether their technology falls within the scope of the law's audit requirements. This discretion can undermine accountability by allowing companies to decide that a technology is out of scope, even if the system is potentially biased and meaningfully impacts employment outcomes.

To ensure that auditing standards lead to high quality audits and provide appropriate assurance about compliant audits, groups have studied the outcomes of requirements for audits in existing laws and synthesized lessons from other auditing domains such as finance and medicine~\cite{raji2022outsider, wright2024null, groves2024auditing, lam2024framework, gerchick2025auditing}. This body of work has collectively identified several key considerations for the design of effective standards. Some groups have called for standards that better account for their users' needs and work contexts~\cite{ojewale2025towards, schor2024mind}. Others have discussed how the design of audit requirements must enable auditors to draw unambiguous conclusions about the audit target and enable audit stakeholders to draw unambiguous conclusions about a completed audit's compliance with the standard. Drawing on their experiences conducting audits under LL144, \citet{lam2024framework} emphasize that requirements must be ``verifiable or observable conditions'' that ``must enable auditors to form an unambiguous opinion about whether a given criterion is satisfied.'' \citet{wright2024null} further emphasizes the importance of unambiguously verifiable criteria through their discussion of how the discretion that the language in LL144's requirements grants to companies makes it difficult for outsiders to evaluate whether the company complies with the law, a determination that drives critical actions such as regulator demands that a company stop using a given technology. Taken together, this line of work highlights the importance of verifiable and observable requirements that support unambiguous conclusions \cite{lam2024framework, wright2024null, raji2022outsider}, requirements that collectively support holistic evaluations \cite{lam2024framework}, and context-specific requirements that are compatible with the practices and contexts of those using the standard \cite{schor2024mind, ojewale2025towards}. We seek to expand these efforts to build an evidence base for the design of auditing standards by studying aspects of the design of auditing standards that undermine its effectiveness.

%% file: 4-methods.tex
\section{Methods}
\label{sec:methods}

\subsection{Data Collection}
\label{sec:methods-data-collection}
\subsubsection{ASB 018 standard}
To understand ASB 018's goals, we drew on both the standard itself and the factsheet that the ASB provides to accompany the standard.

\subsubsection{PGS audits (i.e., internal validation studies)}
To identify existing audit practices, we gathered PGS audit reports from the American Society of Crime Laboratory Directors Validation and Evaluation Repository\footnote{The repository can be accessed at this link: \url{https://www.ascld.org/validation-evaluation-repository/}.}. This database is, to our knowledge, the only publicly available centralized and voluntary repository for U.S. forensic labs' PGS audits. From this repository, we found five PGS audit reports, all for the STRmix PGS: the NYC Office of the Chief Medical Examiner (OCME) Lab's internal validation of STRmix v2.4~\cite{ocmeIV} and STRmix v2.7~\cite{ocmeIV2.7}, the Colorado Bureau of Investigation (CBI) Lab's internal validation of STRmix v2.5~\cite{cbiIV}, the Palm Beach Sheriff’s Office (PBSO) Lab's internal validation of STRmix v2.6.2~\cite{pbsoIV}, and the Maryland State Police's (MSP) internal validation of STRmix v2.9.1~\cite{mspIV}.\footnote{Because the creators of STRmix claim that the software is being used in 91 labs in the U.S.~\cite{strmix2025use}, we expect there to be somewhere on the order of 100 audit reports. It is important to note that at the time of our study, this database only contained five.} This dataset covers a range of software versions, state and city forensic labs of different sizes and geographies, and years (2016-2024).\footnote{The first U.S. forensic lab internal validation of STRmix was completed in 2014~\cite{butler2024history}.} In total, we analyzed over 280 pages of audit reports.

\subsection{Assessing Study Compliance}
\label{sec:methods-assessing-compliance}
The first author first assessed each audit report against relevant ASB 018 requirements to understand whether the audit could be interpreted as complying with the standard. Importantly, only the MSP lab's audit report (completed in 2024)~\cite{mspIV} claims compliance with ASB 018. Three audits were completed before the publication of ASB 018 in 2020 (OCME v2.4~\cite{ocmeIV}, CBI v2.5~\cite{cbiIV}, PBSO v2.6.2~\cite{pbsoIV}). While the OCME lab's v2.7~\cite{ocmeIV2.7} audit report was completed in 2021, it does not mention ASB 018. Thus, our focus is on interpreting whether labs' practices \textit{could be interpreted} as complying with the standard. 

Furthermore, of the three audit reports completed before 2020, none of the labs uploaded updated reports to the repository after 2020, despite ASB 018 advising labs to review their audits and supplementing where necessary~\cite{asb018}. This not only suggests that labs with pre-2020 audits believe their audits to be compliant with ASB 018, but also that ASB 018 did not change labs' existing practices. We further expand on this observation in Section~\ref{sec:disc-compatibility}.

\subsubsection{Identifying line-level requirements of ASB 018} 
\label{subsubsec:line-level-reqs}
Adopting an approach similar to \citet{lawrence2023bureaucratic}, we extracted all of the individual line-level requirements in ASB 018. We interpreted any action proceeded by ``shall'' as a line-level requirement, and focused only on requirements for internal validation studies. Next, we wrote out, as unambiguously as possible, how we interpreted each requirement. This step was crucial due to multiple ambiguities in ASB 018's language. In general, when we identified ambiguity in a requirement, we used the most permissive interpretation. For example, to assess compliance with ASB 018's requirement that labs ``shall evaluate both the appropriate sample types [...] and the number of samples within each type [and] shall base this evaluation on the intended application of the software'' (R4.1.6), we asked: ``Does any part of the report claim that the study's evaluated samples are appropriate for the intended application of the software?''  We provide additional details and the full list of our interpretations of each of the line-level requirements used in our study in Appendix~\ref{sec:appendix-b-interpretations}.

\subsubsection{Assessing study compliance with each of the requirements of ASB 018} 
For each audit, the first author read through the audit report, and resolved any confusions about report language and study design with the fourth author. Then, for each study, the first author identified a portion of the report that could be interpreted as fulfilling our interpretation of each of ASB 018's line-level requirements. 

Throughout this process, ASB 018's design also created challenges for comprehensively assessing compliance. For example, ASB 018 requires that samples studied ``represent (in terms of number of contributors, mixture ratios, and the total DNA [amount]), the range of actual casework samples intended for analysis with the system at the laboratory''~\cite{asb018}. While this requirement is not subject to the ambiguities discussed in Section~\ref{subsubsec:line-level-reqs}, it was difficult for us to comprehensively assess compliance with this requirement because none of the labs explicitly stated their lab's intended range of analysis. Since ASB 018 does not require that labs explicitly state the range of samples intended for analysis, we did not interpret reports' lack of specified ranges as non-compliance. Instead, following our permissive approach to assessing study compliance, we accepted any language in the report that suggests the lab chose samples representative of their intended range (e.g., report language concluding that the audit demonstrated the software was ``fit for its intended purpose in the lab'') as evidence of study compliance with this requirement. Table~\ref{appendix-table-compliance} in Appendix~\ref{appendix-b-compliance} documents sections of each audit report that we took to fulfill each of our interpretations of ASB 018's line-level requirements.

\subsection{Data Analysis}
\label{sec:methods-data-analysis}
Our study investigates the gaps between internal validation practices that ASB 018 envisions and the internal validation practices it enables, and how the design of ASB 018 enables these gaps. To do so, we first identified ASB 018's goals for compliant audits, i.e., what practices ASB 018 envisions or seeks to ensure about compliant studies. Next, we examined how labs have conducted audits in practice. Last, we returned to the standard and examined how its design enables the gaps between the envisioned and realized practices identified in the previous two steps.

We analyzed ASB 018 using five audit stages from \citet{ojewale2025towards}'s framework for stages of an audit: Audit Scope, Standards Identification, Performance Analysis, Audit Judgment, and Audit Communication. We used these stages to connect our study to existing literature and ensure our findings are relevant to other audit contexts, and we focused on this specific subset of stages given their relevance to the PGS context we are studying. For each audit stage, the first author identified and qualitatively coded relevant excerpts from the standard and the factsheet. We also coded these documents to investigate how ASB 018 envisions the purpose of audits. The first author then identified relationships between codes within each audit stage (and also for audit purpose) and grouped codes together into increasingly abstract themes through an iterative, bottom-up affinity diagramming process~\cite{beyer1999contextual}. The group then further refined these themes through discussion, focusing on themes that captured key meanings within each audit stage and assessing the fit of each theme with its corresponding audit stage. We used our themes for ASB 018's vision for audit purpose to complement our construction of themes for the standard's goals for each audit stage. For example, we drew on our interpretation of the standard's emphasis on the role of audits in ``establishing boundaries on software use'' in our interpretation of one of the standard's goals as ``expectations are falsifiable''.

Next, to explore potential gaps between laboratories' validation practices and ASB 018's goals, the first author identified audit excerpts relevant to each of the ASB 018 themes and inductively coded those excerpts. Instead of looking for specific practices, we grounded our interpretations in the practices observable in the report. For each audit stage, the team discussed the initial list of codes to ensure codes captured meaningful divergences from the standard's goals and then iteratively grouped codes into higher-level themes.

Finally, we sought to understand how the design of the standard enables the gaps we identified in the previous step. For each ASB 018 theme, the first author identified relevant ASB 018 requirements. The first author analyzed these excerpts alongside the codes and high-level themes describing internal validation practices, and generated codes capturing specific characteristics of the requirements that enable observed patterns in validation study practices. The first author then iteratively grouped codes into themes.

For example, under the Audit Scope stage, a theme representing ASB 018's goals was ``test broader sociotechnical system'' (corresponding codes included ``internal validation as providing assurance about system'' and ``acknowledging impact of user input''), a theme representing internal validation practices was ``testing software only, not including user'' (codes included ``using ground truth number of contributor value''), and a theme representing aspects of ASB 018's design that enabled the mismatch between goal and practice was ``isolated requirements'' (codes included ``addressing performance measurement and user input separately'').

\subsection{Study Limitations and Future Work}
\label{sec:methods-limits-fw}
\subsubsection{Audit artifacts as proxies for audit practice}
We seek to understand current audit practices by examining how they are discussed and represented in publicly available audit reports. In doing so, our findings our based on accounts of practices. Because these reports are typically submitted in court, we expect them to accurately represent auditing practices. We hope that our findings can compel increased transparency about audit practices so future work can examine audit practices using a variety of sources and methods to investigate the extent to which audit reports authentically and fully capture audit practices.

\subsubsection{Representativeness}
We analyzed audit reports produced by labs of varying types (city, county, and state), geographies, and sizes (e.g. large and small labs), for a variety of versions of the STRmix probabilistic genotyping software. We observed many similarities in the language and structure of the audit reports we examined, and we attribute these similarities to labs' (i) reliance on SWGDAM validation guidelines, a widely used alternative source of guidance for internal validation studies, and (ii) use of a reporting template provided by STRmix developers. These observations suggest that our findings may extend to audits that also reference the SWGDAM guidelines and audits by labs that also use the STRmix software.

However, we caution against assuming that these findings are applicable to all validation studies. At present, while the creators of STRmix claim that it is being used in 91 labs in the US and 29 more internationally~\cite{strmix2025use}, few labs have made their audit reports publicly available, which limits the extent to which we can conclude how representative these reports are~\cite{butler2024dna}. We hope that our findings can support efforts underway that are advocating for greater transparency of audit reports (e.g., \cite{canellas2021defending, butler2024dna, abebe2022adversarial, jin2024beyond, krane2022using, matthews2019right, matthews2020trusted, kinshipproblem}) that would enable future work that can additionally leverage quantitative methods to explore the representativeness of our findings of the broader landscape of audit reports.

\subsubsection{Our permissive approach to assessing study compliance}  
While we drew on our collective familiarity and expertise in PGS and validation studies, our view does not necessarily represent the view of those who might actually be tasked with assessing laboratory audits for compliance with ASB 018, such as organizations and auditors tasked with forensic laboratory oversight and accreditation. Our study takes a permissive approach, towards our goal of identifying what practices are \textit{possible} under ASB 018. This allows us to investigate the largest possible gaps between auditing practices and the standard’s desired practices, to raise awareness of what could possibly happen and to help safeguard against these possibilities. Future work can investigate how actors tasked with forensic laboratory oversight and accreditation interpret ASB 018's requirements and the extent to which their interpretations vary.

\subsubsection{Critically examining ASB 018's goals} 
Our study aims to assess current audit practices against ASB 018's goals for PGS audits, towards our goal of assessing ASB 018's effectiveness and how the design of the standard may undermine its effectiveness at ensuring compliant audits meet its goals. A critical area for future work to investigate, to complement our assessment of the ASB 018's effectiveness, is the degree to which ASB 018's own goals comprehensively assess PGS system effectiveness. 

\begin{figure*}[h!]
    \centering
    \includegraphics[width=\textwidth]{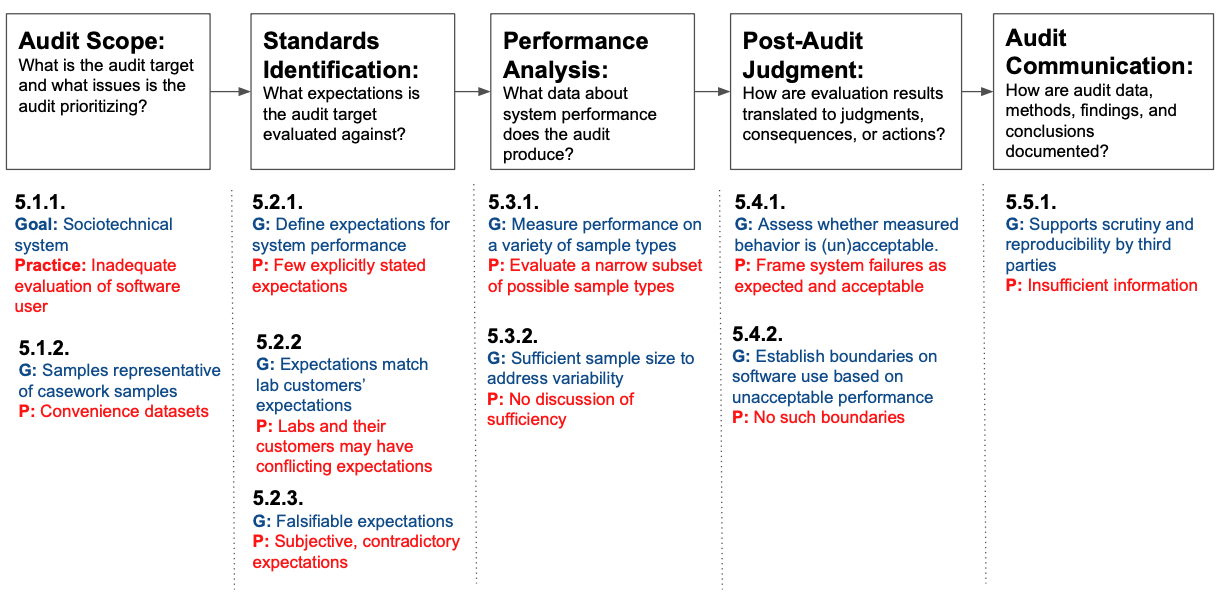}
    \caption{Summary of the gap between each ASB 018 goal (blue text starting with \textbf{G:}) and the corresponding compliant but unsatisfactory audit practices (red text starting with \textbf{P:}). We identify gaps between ASB 018's envisioned practices and real-world audit practices for each of five stages of the audit process, which we draw from ~\citet{ojewale2025towards}.}.
    \label{fig:results-summary}
    \Description{Flowchart with five stages flowing from left to right: Audit Scope, Standards Identification, Performance Analysis, Post-Audit Judgment, and Audit Communication. For each stage, we frame the audit stage as a question that auditors would ask at the audit stage, and then we summarize our findings on the gaps between ASB 018's goals and compliant but unsatisfactory audit practices. Each gap corresponds to a subsubsection of the Results. Question for Audit Scope is: "What is the audit target and what issues is the audit prioritizing?" The two Goal-Practice gaps under Audit Scope are (1) Goal: Sociotechnical system, Practice: Inadequate evaluation of software user; (2) Goal: Samples representative of casework samples, Practice: Convenience Datasets. Question for Standards Identification is: "What expectations is the audit target evaluated against?" The three Goal-Practice gaps under Standards Identification are (1) Goal: Define expectations for system performance, Practice: Few explicitly stated expectations; (2) Goal: Expectations match lab customers’ expectations, Practice: Labs and their customers may have conflicting expectations; and (3) Goal: Falsifiable expectations, Practice: Subjective, contradictory expectations. Question for Performance Analysis is: "What data about system performance does the audit produce?" The two Goal-Practice gaps under Performance Analysis are (1) Goal: Measure performance on a variety of sample types, Practice: Evaluate a narrow subset of possible sample types; and (2) Goal: Sufficient sample size to address variability, Practice: No discussion of sufficiency. Question for Post-Audit Judgment is: "How are evaluation results translated to judgments, consequences, or actions?" The two Goal-Practice Gaps under Post-Audit Judgment are: (1) Goal: Assess whether measured behavior is (un)acceptable, Practice: Frame system failures as expected and acceptable; and (2) Goal: Establish boundaries on software use based on unacceptable performance, Practice: No such boundaries. Question for Audit Communication is "How are audit data, methods, findings, and conclusions documented?" The one Goal-Practice gap under Audit Communication is (1) Goal: Supports scrutiny and reproducibility by third parties, Practice: Insufficient information.}
\end{figure*}

\subsection{Positionality}
All authors were either trained as researchers in or work in the United States. The first author is a researcher in algorithm auditing and human-computer interaction. The second author is a researcher trained in statistics, and the fifth author is a researcher in empirical machine learning who has also served as an AI and law advisor to numerous defense, civic, and government organizations. Collectively, these authors' background and familiarity with auditing literature informed their decision to analyze ASB 018 and audit reports through the lens of five stages of the audit process and shaped the patterns they identified in the standard's design and the auditing practices.  

The third and fourth authors are experts in forensic DNA profiling and PGS, and the third author is also a researcher trained in molecular biology and population genetics: they have written numerous articles on the subject in forensics literature and frequently testify as expert witnesses in criminal cases involving evidence produced by forensic lab use of PGS. Throughout their casework, they have examined dozens of laboratory PGS validation studies and have encountered a variety of laboratory uses of PGS. Our specific focus on ASB 018 is driven by their experiences observing laboratory references to ASB 018 in criminal cases and their perspectives on ASB 018 as a standard with qualities that meaningfully differentiate the standard from other existing PGS validation guidelines in forensic DNA. Their expertise and experiences with investigating real-world uses of PGS in criminal cases also focused our analysis of audit practices on practices they commonly observed that created meaningful risks to defendants subject to evidence produced using these software.

%% file: 5-results.tex
\section{Results: Gaps between ASB 018's goals (envisioned audit practices) and the audit practices it enables}
\label{sec:results}
From our analysis of study compliance with ASB 018, we found that all the audits we examined could be interpreted as compliant with our permissive interpretation of ASB 018's requirements language. 

In this section, we present our findings on the gaps between ASB 018's envisioned audit practices and the audit practices it enables, and how the design of ASB 018 enables these gaps. First, we present our findings on how ASB 018 envisions the \textit{purpose} of audits, since we use this to inform our interpretation of ASB 018's goals for each of the five audit stages. We organize the remainder our findings by the five audit stages outlined in Section~\ref{sec:methods}. In each subsection, we first present each of the goal-practice gaps and then discuss aspects of ASB 018's design that enable these gaps. Figure~\ref{fig:results-summary} summarizes our findings on the goal-practice gaps for each audit stage.

\textbf{Audit Purpose:} Audits help laboratories \textbf{(i) establish boundaries on software use that delineate the range of DNA samples for which the audit has demonstrated acceptable PGS system performance.} These boundaries, in turn, \textbf{(ii) shape protocols for software use in casework} and \textbf{(iii) provide assurance to stakeholders that the likelihood ratio evidence produced using the PGS system in an individual criminal case is reliable.} ASB 018 explicitly states this goal when defining internal validation as the ``acquisition of test data within the laboratory to [...] determin[e] [...] limitations of the system'' ~\cite[p.2]{asb018}. The factsheet provides additional context suggesting that these limitations are boundaries on software use in casework: validation studies ``define limitations for its use'' that ``establish the range of DNA profiles upon which the program may be used effectively''~\cite{asb018factsheet}. Through these statements, we understand  limitations as delineations between the types of DNA profiles for which internal validation studies have demonstrated acceptable PGS system performance that others may ``have confidence in''~\cite{asb018factsheet}, and, the types of DNA profiles for which the software has not been shown to be sufficiently reliable and should therefore not be used. 

\subsection{Audit Scope: Determining the Audit Target and Selecting Issues to Prioritize Investigating in the Audit.}
\label{sec:scope}
\subsubsection{Gap: Broader sociotechnical system vs. Probabilistic genotyping software.}
\label{sec:scope-gap1-target}

\textbf{\\Goal: Compliant audits evaluate the broader sociotechnical system that takes a crime scene sample as input and produces a likelihood ratio (LR) as output. This system encompasses both the PGS and the PGS user.} When using the PGS to produce a LR, a forensic lab analyst specifies key inputs to the software such as their estimate for the \textit{number of contributors (NoC)} to the DNA sample. ASB 018 highlights the ``user input parameters'' as key components of the broader sociotechnical system that audits must ``evaluat[e]''~\cite[R4.1.4]{asb018} and ``address''~\cite[Annex R4.1.4]{asb018}. Testing this broader sociotechnical system is also crucial to ASB 018's end goal for audits to provide assurance about the reliability of LR evidence---a \textit{system} output---presented in a criminal case.

\textbf{Practices:} \textbf{Several audits did not use the user's estimate of the \textit{number of contributors (NoC)} parameter: a key user input that significantly impacts the reliability of the final LR output and is frequently estimated incorrectly.} While the OCME and CBI labs' audit reports describe using user estimates of the NoC when testing software sensitivity and specificity, the PBSO and MSP labs' audit reports describe using the \textit{ground truth} NoC. These audits' use of the ground truth NoC is especially concerning given the frequency of incorrect NoC estimates in the audits that used user NoC estimates and their consequent impacts on the LR outputs. For instance, the OCME v2.7 audit report reveals that users underestimated the NOC for 25 out of 40 five-person samples and noted that several false negative results ``occurred with [five-person DNA mixtures] where the apparent NoC was [underestimated]''~\cite[p.21]{ocmeIV2.7}. 

\textbf{Some audits claimed that software users would be able to prevent observed software failures in casework, despite having no experiments that tested software users' ability to identify and prevent such failures.} For example, the OCME v2.7 audit observed several false negative LRs but claimed that a ``trained analyst'' would be able to notice that the software's intermediate outputs were ``unintuitive,'' re-run the software with a correction, and produce correct results (i.e., true positive LRs)~\cite[p.8]{ocmeIV2.7}. However, without incorporating the analyst's inspection of these intermediate outputs into the testing process, the audit provides no assurance that a user would correctly identify and fix such failures in casework. In other words, the audit report claims that the \textit{system} would produce a correct output when analyzing similar samples in casework but does not conduct tests that would provide evidence to support the claim and provide assurance of system reliability in casework.

\textbf{We also found evidence in several reports that suggested that the PGS developers, and not the labs, conducted parts or all of the audit.} This suggests that any PGS user decisions that were used in the audit may have come from the developers or other company employees, as opposed to the lab's analysts who would be operating the software in casework. For instance, the CBI v2.5 audit report begins with a memo from the lab's quality director stating: ``[ESR] submitted the summary and associated data files for the [audit] for the [laboratory]''~\cite[p.1]{cbiIV}. The company's copyright on the report further suggests developer involvement in the study. While the OCME v2.4 audit report does not explicitly state developer involvement, presence of the ESR copyright also suggests developer involvement~\cite{ocmeIV}. This evidence suggesting that developers may have conducted part or all of these audits raises concerns that even when tests incorporated user estimates, the users in the audit may not have been following the same protocols or training in the same environment as laboratory analysts who would be using the software in casework.

\subsubsection{Gap: Inputs representative of casework samples vs. Inputs chosen out of convenience.}
\label{sec:scope-gap2-inputs}
\textbf{\\Goal: Compliant audits test the PGS system using inputs that represent the range of samples the lab (i) plans to analyze in casework and (ii) will likely encounter in casework.} ASB 018 requires that audits ``include case-type profiles''~\cite[R4.1.3]{asb018}, which the standard defines as samples ``exhibiting features that are representative of a plausible range of casework conditions [such as] [...] shared alleles''~\cite[p.1]{asb018}. ASB 018 also expects that these profiles ``represent [...] the range of actual casework samples intended for analysis with the system at the laboratory''~\cite[R4.1.3]{asb018}.

\textbf{Practices:} \textbf{Audits primarily constructed test samples out of convenience---not out of a formal assessment of samples the lab intends to analyze or is likely to encounter in casework.} Labs frequently encounter mixtures containing DNA from related individuals in casework, and analysis of these mixtures is more likely to produce false positives due to individuals having similar DNA profiles (i.e., sharing alleles) \cite{kelly2022exploring}. While the CBI v2.5 and MSP v2.9.1 audits constructed mixtures specifically to test scenarios where contributors to the mixture are relatives~\cite{cbiIV, mspIV}, the PBSO v2.6.2 audit did not explicitly test system performance on mixtures with related contributors and instead described the mixtures in their test set as having ``varying amounts of allele sharing''~\cite[p.10]{pbsoIV}. The OCME v2.4 and v2.7 audit reports use similar language~\cite{ocmeIV, ocmeIV2.7}. This language suggest that these audits did not design their test set to account for mixtures of related individuals, especially since any set of mixtures is likely to have some amount of allele sharing (i.e., DNA profiles of any two unrelated individuals are likely to contain some of the same alleles).

\subsubsection{How ASB 018's design enables these gaps}
\label{sec:scope-design}
\textbf{ASB 018's \textit{narrow definition} of the audit target misaligns with its vision that audits provide assurance of system reliability in casework.} ASB 018 acknowledges the importance of incorporating the software user's decisions into audits and implies a broad audit scope when describing audits as providing assurance of the reliability of the LR output in criminal cases. However, ASB 018 defines ``probabilistic genotyping system'' as the ``software, or software and hardware''~\cite[p.2]{asb018}, failing to acknowledge the lab analyst whose use of the PGS shapes the ultimate LR output. 

\textbf{The organization and language of ASB 018's requirements treat performance measurement and user inputs as separate items for audits to ``include.''} ASB 018 first requires that audits ``address [...] accuracy, sensitivity, specificity, and precision''~\cite[R4.1.3]{asb018}. In a separate requirement, ASB 018 requires that audits also ``include evaluati[ons of] user input parameters''~\cite[R4.1.4]{asb018}.  Instead of requiring that labs account for and measure how user inputs impact system accuracy, sensitivity, specificity, and precision, ASB 018 treats these two components as two separate criteria that audits can address in isolation of one another. For instance, labs can satisfy the first requirement by measuring PGS performance without considering user inputs and the second requirement with a separate study that changes user input parameters and observes whether LRs increase or decrease \textit{without} measuring performance~\cite{pbsoIV, mspIV}.

\textbf{ASB 018 uses vague language that allows labs to do very little to comply.} As discussed earlier, ASB 018 requires that audits ``shall include evaluating user input parameters,'' but does not specify what this evaluation must entail~\cite[R4.1.4]{asb018}. ASB 018's requirement for labs to ``consider the effect of overestimating and underestimating the number of contributors'' is similarly vague. There is no requirement to measure the impact of user inputs on system accuracy, sensitivity, specificity, and precision.

\textbf{ASB 018 envisions that labs conduct audits alone, but its use of vague language and failure to consider how developers might conduct part or all of an audit does not prohibit audits conducted by developers.} ASB 018 frames most requirements at the level of the internal validation study without specifying the actor responsible for carrying out the required action. When ASB 018 does address the lab, it does not clearly specify which audit components the lab is responsible for. Take, for example, the requirement that ``the laboratory shall validate [the] PGS system''~\cite[R4.1]{asb018}. Does a lab's review of a summary of the audit conducted by the developers count as the lab ``validating'' the software? Can the developers provide the inputs to the software using their estimates for user inputs such as the number of contributors?

\textbf{ASB 018 does not require that labs take note of the scenarios they typically encounter or intend to analyze---a preliminary step that would hold labs accountable to some specific range of samples.} ASB 018 requires that audits test ``case-type profiles [...] that represent [...] the range of actual casework samples intended for analysis''~\cite[R4.1.3]{asb018}, but does not require labs to determine and specify this range---a preliminary step that would help hold labs accountable for creating test sets that represent this range of intended uses.

\subsection{Standards Identification: Articulating the Criteria or Expectations the System is Held to in the Audit}
\label{sec:stds}

\subsubsection{Gap: Predetermine specifications that define acceptable system performance vs. Failure to explicitly define specifications}
\label{sec:stds-gap1}
\textbf{\\Goal: Labs test the PGS system against predetermined specifications: expectations for acceptable system performance.} In its definition of ``validation,'' ASB 018 emphasizes the importance of ensuring the PGS system ``will consistently [meet] its predetermined specifications''~\cite[p.2]{asb018}. Taken together with Requirement 4.1.3 and ASB 018's vision for audit purpose, we understand ASB 018 to further suggest that these specifications must define acceptable ``accuracy, sensitivity, specificity, and precision'' of the PGS system's LR output~\cite[R4.1.3]{asb018}.

\textbf{Practices: Audit reports often do not explicitly define expectations for acceptable software behavior, suggesting that audits were not evaluating the system against predetermined expectations for acceptable performance.} Instead of explicitly stating the expectations they were evaluating against, audits sometimes implicitly communicated their expectations by describing observed results as ``expected.'' For example, after presenting observed true positive rates, the MSP v2.9.1 audit report concludes: ``As expected, there was a high percentage of [true positive] LRs''~\cite[p.27]{mspIV}. This language suggests that the lab may have understood a ``high'' true positive rate to be acceptable system sensitivity, but the audit never explicitly stated this expectation. At other times, audits simply concluded by presenting results without any language suggesting that results met expectations, making it difficult to infer what the lab's expectations were or whether they had \textit{any} expectations to begin with. For example, the OCME v2.4 audit measured software precision (i.e., the degree of variability in LRs between multiple software runs on the same inputs) and concluded that the observed LRs ``demonstrate that [...] the values obtained for the various runs remain close; within one order of magnitude except for the four-person mixture''~\cite[p.19]{ocmeIV}.

\subsubsection{Gap: Expectations align with forensic labs' customers' expectations for PGS performance vs. Expectations that may conflict with customers' desires}
\label{sec:stds-gap2}
\textbf{\\Goal: Compliant audits test the system against expectations for acceptable system performance that align with lab customers' perceptions of what constitutes acceptable system performance.} ASB 018 emphasizes that ``[audits] provide the study results and conclusions necessary for customers of [forensic labs] to have confidence in the [LR] evidence provided''~\cite[Foreword]{asb018}. In order for audits to support customer\footnote{We interpret ASB 018's use of the word ``customer'' to imply stakeholders who pay laboratories for their services (e.g., prosecutors). Importantly, ASB 018 does not use the word ``stakeholder'', which refers to a broader set of actors.} assurance of PGS system reliability, the audit must assess PGS system behavior against lab customers'  expectations (as opposed to the PGS developer or forensic lab's expectations).

\textbf{Practices:} \textbf{Audit reports suggest that audits relied on expectations for acceptable system performance that may misalign with customers' needs, and none of the audit reports mention customers' needs or perceptions.} For example, the MSP v2.9.1 audit report concludes that the audit's observed false positive rates of up to 1.3\% and false negative rates of up to 16\% demonstrate that the PGS performed ``as expected''~\cite[p.27]{mspIV}. In doing so, the audit did not consider how a defense attorney's or a prosecutor's perception of acceptable false positive or false negative rates may differ. Furthermore, we found that none of the audits even mentioned customers' needs or perceptions in their reports, suggesting that labs did not consider whether their expectations aligned with those of their customers.

\subsubsection{Gap: Falsifiable expectations vs. Subjective, otherwise unclear, and contradictory expectations.}
\label{sec:stds-gap3}
\textbf{\\Goal: } \textbf{Expectations for acceptable system performance are falsifiable.} In other words, labs define what constitutes acceptable system behavior using criteria that enable labs to assess observed system behavior against these criteria and unambiguously state whether the system meets these criteria. Crucially, labs must be able to unambiguously identify when software behavior \textit{fails} to meet these criteria for acceptable performance, since the end goal of an audit is to establish boundaries on acceptable software use.

\textbf{Practices: }\textbf{Audits described expectations using ambiguous criteria.} For instance, audits frequently defined expectations using subjective criteria such as ``high'' and ``low.'' For example, all audits stated that the LR for true contributors ``should be high.'' The MSP v2.9.1 audit report concludes that their observed sensitivity rates demonstrated ``high [PGS] sensitivity''~\cite[p.23]{mspIV}. Audit reports also use other subjective criteria when discussing expectations, such as ``intuitive'' (e.g., ``result in an intuitive inclusionary LR''~\cite[p.7]{ocmeIV2.7}), ``generally'' (e.g., ``true contributors generally gave high LRs''~\cite[p.21]{ocmeIV2.7}), and ``close'' (e.g., ``the values obtained for the various runs remain close''~\cite[p.19]{ocmeIV}). \textbf{Audits also used otherwise unclear language in their expectations.} All audits' definitions of sensitivity imply that they expected the PGS to ``reliably resolve the DNA profile of [true] contributors''~\cite{ocmeIV, ocmeIV2.7, pbsoIV, mspIV, cbiIV}. However, none of the audit reports specify what it means for a PGS to ``resolve'' a DNA profile. For example, does an LR above 1 count as ``resolving'' the profile, or should the LR be above some other threshold value like 1,000? \textbf{Expectations within the same report were sometimes contradictory.} For example, the OCME v2.4 audit report defines specificity as ``the ability of the software to reliably exclude non-contributors''~\cite[p.6]{ocmeIV}. However, when discussing experiment results, the report states that some false positive LRs were ``expected''~\cite[p.14]{ocmeIV}.

\subsubsection{How ASB 018's design enables these gaps}
\label{sec:stds-design}
While ASB 018 refers to ``predetermined specifications'' in its definition of validation, \textbf{the standard does not include any requirement that labs actually \textit{define} these specifications}. Furthermore, despite emphasizing the importance of audits establishing boundaries on acceptable software use, \textbf{ASB 018 does not establish any requirements for \textit{how} these specifications should be defined to ensure that defined expectations actually support the end goal to establish boundaries on acceptable software use}. Instead, ASB 018 simply requires that labs ``\textit{address} [...] accuracy, sensitivity, specificity, and precision''~\cite[emph. added]{asb018}. Similar to our discussion of ASB 018's use of the word ``consider'' in Section~\ref{sec:scope-design}, \textbf{the word ``address'' allows labs to comply without clearly and explicitly defining what constitutes acceptable accuracy, sensitivity, specificity, and precision and then assessing PGS system performance against these predetermined specifications}.

\subsection{Performance Analysis: The Actual Evaluation Itself (Gathering Data on System Performance)}
\label{sec:performance}

\subsubsection{Gap: Measure system performance on a variety of sample types vs. Measure performance on a narrow subset of samples tested.} 
\label{sec:performance-gap1}
\textbf{\\Goal: Labs measure sensitivity, specificity, precision of the PGS system's LR outputs for a variety of sample types, and labs measure accuracy for a smaller set of sample types.}
A goal that follows from \textit{defining} expectations for acceptable PGS system performance with respect to sensitivity, specificity, precision, and accuracy is \textit{measuring} these dimensions of system performance. Furthermore, ASB 018 envisions that labs generate these measurements for a variety of DNA sample types---a detail crucial to ensuring that audits help establish boundaries on software use. ASB 018 also envisions that labs measure system accuracy for a variety of sample types but explicitly limits its expectations for labs to a smaller set of sample types: ``[one-person] samples [and simple][two]-person mixtures''~\cite[p.1]{asb018}. ASB 018 emphasizes that situations where ``the ground truth is not known are not suitable for accuracy studies''~\cite[p.1]{asb018}

\textbf{Practices: Audits often measured LR sensitivity, specificity, precision, and accuracy on a narrow subset of the samples that labs had access to or could have tested.} Every report documents audit measurements of software sensitivity and specificity in an experiment specifically dedicated to sensitivity and specificity. \textbf{In these experiments, most audits did \textit{not} explicitly test DNA sample types that labs often encounter in casework, such as significantly degraded\footnote{E.g., DNA samples degrade after exposure to sunlight for several days} mixtures}. Instead, reports only address degraded samples when describing later experiments that often did not measure system specificity, precision, and accuracy. For example, the OCME v2.4 audit artificially degraded a single one-person DNA sample and only investigated the impact of degradation on the true contributor's LR (instead of also investigating non-contributor LRs to test specificity)~\cite{ocmeIV}. \textbf{All audits investigated system precision in separate experiments, and most audits' precision experiment test sets were  \textit{significantly smaller subsets} of their sensitivity and specificity experiment test sets.} Lastly, while ASB 018 explicitly constrains the set of sample types for which labs should measure system accuracy, ASB 018 \textit{does} include simple two-person mixtures in its set of suitable sample types. However, \textbf{all audits' accuracy experiments only evaluated one-person samples.} Audits sometimes even acknowledged that the samples they were testing were a subset of the entire set of sample types they could have tested, stating something like: ``There is a small subset of profiles where [system accuracy can be tested]. These include [one-person samples]''~\cite[p.3]{ocmeIV}. In other words, these audits measured accuracy on one type of sample (which is also one of the types of DNA samples least likely to be associated with poor PGS performance) when they could have also included two-person mixtures.

\subsubsection{Gap: Robust audit measurements vs. Inadequate consideration of sample size sufficiency} 
\label{sec:performance-gap2}
\textbf{\\Goal: Labs produce \textit{robust} measurements of system performance through repeated testing to account for variability in measurements.} ASB 018 recognizes that performance measurements may vary due to upstream DNA preparation procedures, stochasticity in the software's underlying statistical models, and differences between DNA samples. As such, the standard highlights the importance of ``repeated testing'' on a ``sufficient'' range of sample types and number of samples of each type~\cite[Annex A, R4.1.3]{asb018}.

\textbf{Practices:} \textbf{Audits neither commented on the sufficiency of the sample sizes used to test system performance, nor provided measures used to determine whether sample sizes were sufficient.} This lack of discussion is especially concerning since sample sizes varied drastically between audits. For example, when testing system precision, the MSP v2.9.1 audit used one four-person sample~\cite[p.65]{mspIV}, whereas the OCME v2.7 audit used eight samples spanning two- to five- contributor mixtures~\cite[p.28]{ocmeIV2.7}. For sensitivity and specificity, while the MSP v2.9.1 audit tested 30 two-person and 17 five-person mixtures~\cite[p.14-15]{mspIV}, the OCME v2.7 audit tested 169 two-person and 40 five-person mixtures~\cite[p.15]{ocmeIV2.7}.

\subsubsection{How ASB 018's design enables these gaps}
\label{sec:performance-design}
\textbf{ASB 018's vague language and use of separate line-level requirements to address interrelated concepts allows labs to analyze PGS performance on narrow subsets of test sets.} While ASB 018 requires that internal validation studies to address ``accuracy, sensitivity, specificity, and precision'' in one line-level requirement, another line-level requirement states, ``These studies shall include [samples] that represent [....] the range of actual casework samples intended for analysis''~\cite[R4.1.3]{asb018}. In this second requirement, ASB 018 does not specify whether ``[t]hese studies'' broadly refers to audits or narrowly refers to individual accuracy, sensitivity, specificity, and precision experiments within these audits (which the standard also refers to as ``studies'' elsewhere, e.g., ``accuracy studies''~\cite[p.1]{asb018}). A permissive interpretation of the second requirement would interpret ``[t]hese studies'' to mean ``[i]nternal validation studies,'' which would allow for a precision experiment to test a narrow subset of test samples while the audit as a whole satisfies the requirement for audits to test representative DNA samples. Furthermore, as we have discussed earlier (Section~\ref{sec:scope-design}), ASB 018 treats complex sample characteristics such as degradation and software performance measurement as two separate criteria that audits can address in isolation from one another.

\textbf{ASB 018 lists only three dimensions of DNA sample types in the requirements, failing to capture key dimensions that the standard outlines in its Terms and Definitions section.} The standard requires that audits include case-type profiles ``that represent \textit{(in terms of number of contributors, mixture ratios, and total DNA [amount])} the range of actual casework samples intended for analysis''~\cite[R4.1.3, emph. added]{asb018}. Here, the standard names three DNA sample characteristics that labs must account for: the number of contributors, mixture ratios, and the total DNA amount. However, the standard's definition of ``case-type profiles'' names other dimensions to consider, such as the amount of ``degradation'' and the number of ``shared alleles'' between individual contributors in a mixture~\cite[p.1]{asb018}. ASB 018's omission of key factors like degradation and allele sharing not only limits attention to the dimensions specified in the requirements, but also introduces ambiguity around whether the dimensions mentioned in the Terms and Definitions section are requirements or suggestions.

\textbf{ASB 018 neither provides criteria for unambiguously determining whether the sample size of a lab's test set is sufficient, nor requires labs to establish their own criteria.} Instead of specifying a minimum number that labs must comply with, ASB 018 requires that labs ``shall perform sufficient studies''~\cite[Annex A, R4.1.3]{asb018}. However, the standard neither specifies how labs should operationalize ``sufficient,'' nor requires labs to establish their own criteria for determining sufficiency. While the standard's requirements language provides labs with the flexibility to design studies that fit their needs and contexts, it also enables labs to test any types and number of samples (e.g., two samples) without justifying why their choice is sufficient. In other words, the standard does not hold labs accountable to any definition of ``sufficient.''

\subsection{Post-Audit Judgment: Discussing Results and Translating Results Into Actions or Consequences}
\label{sec:judgement}

\subsubsection{Gap: Conclude whether measured behavior is acceptable vs. Frame software failures as correct behavior.} 
\label{sec:judgement-gap1}

\textbf{\\Goal: Labs assess observed behavior and determine whether observed behavior is acceptable or unacceptable.} \textit{Providing judgment} about observed system behaviors, rather than stopping at simply producing these observations (i.e., measuring system behaviors), is crucial to ensuring that audits establish boundaries on acceptable software use in casework. In other words, the lab must label a test result as acceptable or unacceptable, instead of simply stating the test result.

\textbf{Practices:  Audits frequently framed software failures as correct behavior. Most audits did not use the word ``failure'' to describe incorrect results, false positives, and false negatives.} For instance, the OCME v2.4 audit report frequently reframes observed false positive results as correct behavior. The audit report states that the false positives were ``not a failure of STRmix''~\cite[p. 13]{ocmeIV}. The OCME v2.7 audit report uses similar language~\cite[p. 24]{ocmeIV2.7} and also frames other software failures as ``non-intuitive,'' stating that some test runs ``resulted in non-intuitive results with an LR = 0 or failure to run''~\cite[p.14]{ocmeIV2.7}. Audits also frequently framed software failures as ``not unexpected'' instead of as failures, responding to false positive and false negative LRs with statements like ``these results are not unexpected''~\cite[p.24]{ocmeIV2.7}. None of the audit reports analyzed in our study frame observed failures as ``unacceptable.''

\subsubsection{Gap: Establish boundaries on acceptable software use in casework vs. Not establishing such boundaries}
\label{sec:judgement-gap2}
\textbf{\\Goal: Labs use their judgments of observed system behavior to establish boundaries on acceptable software use in casework.} As we described at the beginning of Section~\ref{sec:results}, ASB 018 envisions that audits ultimately establish boundaries on acceptable software use, and these boundaries must be specified in terms of DNA sample types.

\textbf{Practices: }\textbf{None of the audits used observed software failures to define boundaries on software use in terms of DNA sample types.} Some audits established another type of boundary: an ``uninformative'' range for LRs (e.g., 0.001 - 1000~\cite[p.24]{ocmeIV2.7}) that the lab would use in criminal cases to present LRs falling in this range as ``inconclusive'', i.e., neither inculpatory nor exculpatory. However, this boundary is not a \textit{restriction} on software use and does not prevent labs from introducing false positive LRs as evidence in criminal cases when the PGS system produces extreme false positive LRs (which is common in the case of marginal samples where there is a high degree of allele sharing between a true contributor and a non-contributor).

\textbf{Oftentimes, in place of boundaries on software use, audits emphasized that analysts would be able to identify the issue and adjust their software use accordingly}. For example, upon observing false positive results, one audit concluded that such scenarios are ``recognizable by trained analysts'' who would be able to correct the analysis accordingly to avoid such errors~\cite[p.8]{ocmeIV2.7}. However, as we discussed in Section~\ref{sec:scope-gap1-target}, audits were scoped around the PGS instead of the broader sociotechnical system, so these audits do not actually provide assurance that analysts would be able to avoid such errors in casework. \textbf{Audits also sometimes concluded without providing any assurance of reliability in casework, instead advising analysts to proceed with caution.} For example, after observing several false negative results, the CBI v2.5 audit simply advised analysts to ``exhibit caution [...] for situations such as this''~\cite[p.31]{cbiIV}. 

\subsubsection{How ASB 018's design enables these gaps}
\label{sec:judgement-design}
\textbf{ASB 018 neither clearly defines ``specification'', nor requires labs to explicitly label whether observed performance is acceptable.} In Section~\ref{sec:judgement-gap1}, we discussed how labs framed software failures as ``not unexpected,'' instead of labeling these failures as unacceptable behavior that the lab should prevent in casework through establishing boundaries on acceptable software use based on audit results. These practices further reveal consequences of the standard's failure to specify what constitutes a ``specification'' (see Section~\ref{sec:stds-design}), and also reveal that the standard fails to ensure that labs actually judge whether audit results are acceptable---a crucial step towards establishing boundaries on acceptable software use.

\textbf{ASB 018 does not explicitly require that audits establish boundaries on software use. Instead, ASB 018 relies on the word ``limitations'', which the standard does not define.} Despite repeatedly describing that audits help determine limitations, ASB 018 does not define what constitutes a limitation. The accompanying factsheet clarifies that limitations are boundaries that define ``the range of DNA profiles upon which the program may be used effectively''~\cite{asb018factsheet}. However, these details are crucially lacking from the standard, the sole document that studies would be assessed against for compliance. Furthermore, \textbf{despite emphasizing that audits establish ``limitations'', ASB 018 does not require a lab to actually \textit{specify} the limitations they identify in the audit.} In other words, the standard seeks to guarantee that labs use audits to determine system limitations, but makes no requirement that labs actually do so.

\subsection{Audit Communication: Documenting Audit Design, Methodology, and Results} 
\label{sec:communication}

\subsubsection{Gap: Documentation enables third party scrutiny and reproducibility vs. Documentation contains insufficient information about test samples and results}
\label{sec:communication-design-gap}\textbf{ \\Goal: Lab documentation of audit data, methods, and results enables third party scrutiny.} 
ASB 018 emphasizes the importance of audit documentation that addresses the information needs of third parties, stating: ``it is incumbent upon any laboratory performing these [audits] to retain these results for the examination and evaluation by third parties. The results should be documented in such a way that the [...] [audits] can be reproduced and decisions made on the basis of these studies documented''~\cite[p.5]{asb018}.

\textbf{Practices: Audit reports often do not provide sufficient information about test samples and conditions that third parties could use to interpret study results.} The audit reports we analyzed omit crucial information about the range of DNA sample types examined, often describing test samples using vague language. For example, each audit report describes their sensitivity and specificity experiment test samples as ``varying'' in their ``[total] DNA [amounts,] mixture proportions'', and ``amounts of allele sharing'' (e.g.,~\cite[p.10]{pbsoIV}). When audit reports do provide specific ranges of sample types explored, they often only provide details for a small set of DNA sample characteristics (e.g., specifying the range of total DNA amounts studied, but not specifying the range of mixture proportions studied~\cite[p.15]{ocmeIV2.7}). \textbf{Audit reports also often do not provide data that would enable third party verification of their claims.} For example, the OCME v2.4 audit report claims that ``The results of all comparisons was expected. [...] the [non-contributors got] very low or 0 LRs''~\cite[p.22]{ocmeIV}. In the appendix, the audit report documents the number of non-contributors with LRs = 0 and plots the LRs in scatterplots, but does not provide exact LR values for non-contributors, despite the plots revealing several false positive LRs~\cite[p.47-50]{ocmeIV}.

\subsubsection{How ASB 018's design enables these gaps}
\label{sec:communication-design}
\textbf{ASB 018 requires that labs document their audits, but non-mandatory wording to describe the importance of ensuring that documentation is accessible and useful to third parties}. ASB 018 requires documentation by stating that ``All [audits] shall be documented and retained by the laboratory'', but switches to using the phrase ``it is incumbent upon'' when emphasizing that labs should document the results to support ``examination and evaluation by third parties''~\cite[Annex A R4.5]{asb018}. \textbf{ASB 018 also fails to specify specific pieces of information that labs must document,} even though the standard specifies in the line immediately after that ``Laboratories shall have a \textbf{summary statement of the sample types} of which the developer used to for their developmental validation''~\cite[Annex A R4.5]{asb018} While the standard could specify a similar requirement that labs document the sample types they tested in their audit, the standard does not include any documentation requirement with this level of specificity. 

%% file: 6-discussion.tex
\section{Discussion}
\label{sec:disc}
Throughout our analysis of ASB 018 and five audit reports, we found that all five audits can be interpreted as compliant with ASB 018 despite falling short of the standard’s goals. We identified gaps between ASB 018’s envisioned practices and the practices it enables across five audit stages and highlighted how the design of the standard enables these gaps.

First, we join \citet{berman2024scoping} in emphasizing the importance of tool developers providing clear definitions of the outcomes they intend for their tool to create. Our qualitative coding of ASB 018 and its accompanying factsheet to develop our interpretations of the goals of the tool’s creators was crucial to our evaluation of ASB 018 because the ASB did not explicitly state the standard’s goals. Without clear statements of intended outcomes, how can one conclude that a tool is effective? Furthermore, in situations like ours where tool developers do not clearly state goals, we contend that clearly articulating perceived goals from the outside, using methods such as our interpretive analysis of the tool’s goals, can help initiate conversations about desired outcomes and compel developers of audit tools to clarify their intended goals, towards holding tool creators (e.g., standards developers) accountable for the tools (e.g., standards) they create and promote. 

We additionally make the following recommendations, which we expand on in the sections below:
\begin{itemize}
    \item Designers of audit standards and other forms of audit tooling should carefully consider how designing a tool for compatibility with users' current perspectives and practices can undermine the tool's effectiveness. This is especially crucial for audit standards where ineffective standards can further legitimize existing practices. (Section \ref{sec:disc-compatibility})
    \item Evaluating an audit standard's effectiveness against fine-grained articulations of the standard's desired outcomes can help identify where increased specificity is crucial for achieving desired outcomes. Existing audit frameworks (e.g., \cite{ojewale2025towards, raji2020closing}) can help increase the specificity with which we articulate these desired outcomes. (Section \ref{sec:disc-design-artic})
    \item Components of audit standards where increased specificity may be crucial to achieving desired outcomes are: (i) actors and actor responsibilities, (ii) minimum activities, and (iii) intermediary steps. (Section \ref{sec:disc-design-specificity})
    \item Co-designing audit standards with groups beyond standards bodies and users of standards is crucial for counterbalancing users' desires for flexibility and can encourage more robust definitions of an audit standard's goals. (Section \ref{sec:disc-design-codesign})
\end{itemize}

\subsection{Carefully consider how designing a tool for compatibility with tool users’ current perspectives and practices can undermine the tool’s effectiveness.}
\label{sec:disc-compatibility}
Prior work in human-computer interaction and algorithm auditing has called for the design of standards that more closely align with the needs and perspectives of those using the standard~\cite{schor2024mind, ojewale2025towards}. For instance, \citet{schor2024mind} highlight a gap between an algorithmic system transparency standard and system designers’ understandings of transparency. They call for standards bodies to more closely align standards with designers' current practices and understandings of transparency. These calls echo recent observations by \citet{berman2024scoping} that much of existing RAI tool development and evaluation prioritizes tool usability, potentially at the cost of effectiveness.

While designing standards to align with users’ needs, workflows, and work contexts can be crucial for encouraging tool use and achieving intended outcomes~\cite{madaio2020co, holstein2019improving, rakova2021responsible}, \textbf{our findings caution that designing a standard to enable flexibility and compatibility with users’ current perspectives and practices can simultaneously undermine the standard’s effectiveness}. While vague language supports user discretion and improves the standard’s compatibility with labs’ existing approaches, it does not hold labs accountable for carrying out critical auditing practices. For instance, while ASB 018's requirement that labs  ``consider the impact of over- and underestimation of the number of contributors'' is compatible with labs' existing practices of observing whether LRs increase or decrease when analysts over- and underestimate the number of contributors to the mixture, it fails to hold labs accountable for measuring how inaccurate estimations of the number of contributors impacts system accuracy, sensitivity, specificity, and precision (Section~\ref{sec:scope-design}).

We contend that \textbf{identifying how a standard’s compatibility may undermine its effectiveness is \textit{especially} critical in the context of standards for algorithm auditing, where ineffective standards not only may fail to improve practices, but at the same time may further legitimize  existing practices.} Three of the five audit reports we examined were completed before the publication of ASB 018 and one audit report published afterwards does not seem to engage with ASB 018, yet all four could be interpreted to comply with the standard’s requirements, suggesting that ASB 018 is compatible with existing lab practices: these labs would not need to significantly change their existing practices to comply with ASB 018’s requirements. In the context of ASB 018, this raises the question: Was the standard designed to improve practices or, as \cite{morrison2020vacuous} warn, was it designed to provide post hoc legitimacy to existing practices? In the next section, we discuss strategies that future efforts to design effective audit standards may leverage to balance flexibility and compatibility with effectiveness and accountability.

\subsection{Recommendations for designing audit standards that effectively support accountability}
\label{sec:disc-design}
\subsubsection{Fine-grained articulation of the audit standard’s desired outcomes and evaluating the standard’s effectiveness at achieving those outcomes can help identify areas where increased specificity is crucial for achieving desired outcomes.}
\label{sec:disc-design-artic}
Prior work on standards and algorithm auditing has framed increased specificity in standards as a double-edged sword: increased specificity can improve audit quality by mandating specific actions and outcomes, but can hinder auditor discretion that may be crucial to make context-appropriate, situational judgments~\cite{raji2022outsider, costanza2022audits, jackson2015standards}. Based on our findings, we suggest that \textbf{fine-grained articulation of outcomes desired by standards bodies and evaluating the standard’s effectiveness using these intended outcomes can help locate specific components where increased specificity is crucial for achieving the standard’s intended outcomes}. For instance, whereas ASB 018's goal for audits to establish boundaries on software use does not demand that labs use a specific ASB 018-mandated metric or threshold to define acceptable performance, it \textit{does} require that labs specify some definition of what constitutes acceptable performance and that these specifications are falsifiable (Section~\ref{sec:stds-design}) -- forms of specificity that ASB 018 crucially omits. Similarly, whereas ASB 018's goal for audits to test samples representative of the range of scenarios that the lab typically encounters or intends to analyze does not demand that all labs test up to five-person mixtures, it does require that labs specify a range that they and others could use to hold them accountable for.

In the context of audit tooling, we recommend using existing auditing frameworks (e.g., \cite{ojewale2025towards, raji2020closing, abebe2022adversarial, lam2024framework, radiya2023sociotechnical}) to define such goals. \textbf{Using \citet{ojewale2025towards}'s framework helped increase the specificity with which we articulated ASB 018’s goals and examined audit practices.} For instance, instead of articulating only the desired outcomes with respect to audit outputs (e.g., establish boundaries), we were able to assess audit reports against practices the standard’s envisions for each audit stage (e.g., scoping the audit, defining performance expectations, and audit documentation). \textbf{This also allowed us to address stages of the audit process beyond evaluation, such as standards identification and audit communication, that are crucial for creating accountability}~\cite{ojewale2025towards}. Echoing calls by \citet{ojewale2025towards} for tools that support all stages of the audit process, we argue that an audit standard’s goals should be articulated with sufficient granularity to address practices throughout the audit process.

\subsubsection{Examine which actors, actor responsibilities, activities, and intermediary steps must be explicitly specified to accomplish the goals of the standard.}
\label{sec:disc-design-specificity}
Recent work studying algorithm audits in hiring has called for increased specificity in how rules for auditing (e.g., audit standards, requirements in audit legislation) define which algorithmic systems are considered in-scope~\cite{wright2024null, groves2024auditing} and who counts as an independent auditor~\cite{groves2024auditing}. Building on this work, \textbf{our findings suggest three components of audit standards where increased specificity may be crucial to ensuring audit standards produce standards developers' desired outcomes and help create accountability: (i) actors and actor responsibilities, (ii) bare minimum practices, and (iii) intermediary steps.}

First, we found evidence that audit responsibilities were distributed between forensic labs and PGS developers in ways that undermined ASB 018’s goals (Section~\ref{sec:scope-gap1-target}). Yet ASB 018’s requirements do not acknowledge and engage with this reality. Many of the standard's requirements were directed at ``internal validation studies'' (as opposed to specific actors), and the few requirements assigned explicitly to labs contained vague language that allows a wide range of lab involvement in audits. This lack of specificity with respect to actors and actor responsibilities not only undermines the effectiveness of the standard, but also risks papering over distributions of auditing responsibilities that raise additional concerns such as lack of auditor independence. Recent work by \citet{groves2024auditing} similarly calls for increased specificity with respect to various actors involved in auditing and their specific roles and responsibilities, highlighting that this specificity can help ease the relational work required of auditors who must navigate tensions between software developers, software users, and regulators. 
Based on our findings, we add that assigning explicit roles and responsibilities to specific actors can not only ensure desired audit practices, but can also make visible the various actors and their roles in shaping audits in ways that open auditing practices to assessment a wider group of stakeholders.

Second, we found ASB 018’s use of vague requirements language such as ``consider'' and ``address'' consistently allowed for audit practices that fell short of the standard’s desired outcomes. For instance, the standard requires that labs ``address [...] precision'', and we see that labs analyzed a small subset of the test set used in the audit’s sensitivity and specificity experiment (Section~\ref{sec:performance-gap1}). In fact, this vague language allows for even more cursory practices. For instance, in PBSO's audit of STRmix v2.6.2, instead of measuring the precision of the software version being validated, the lab suggested that their previous validation of STRmix v2.4 and their sensitivity and specificity studies were sufficient. \textbf{We recommend that designers of audit standards carefully assess the bare minimum an audit could do to fulfill a requirement, and reflect on what specific practices they seek to require instead of relying on vague language.}

Lastly, we found that labs omitted crucial steps necessary to achieve ASB 018’s goals, and ASB 018 enabled this through its failure to require those preliminary steps. For instance, ASB 018's requirement that ``[audits] shall include [samples] that represent [...] the range of actual casework samples intended for analysis with the system at the laboratory'' depends on labs first determining and specifying the range of samples they intend to analyze with the software, but ASB 018 does not require that labs do so (Section~\ref{sec:scope-design}). These omissions also limited our ability to more thoroughly confirm study compliance (Section~\ref{sec:methods}). For instance, similar to how \citet{wright2024null}'s ``null compliance'' signals a lack of publicly available information needed to make a concrete determination of compliance, we found that labs did not provide a specific range of DNA samples that we could use to concretely assess the range of DNA samples tested in the audit. Consequently, following our permissive approach to assessing compliance, we simply accepted a lab's claim that their test set covers the lab's intended range. Requiring these preliminary steps can ensure  that those assessing audit compliance with a standard have crucial information needed to make unambiguous conclusions about audit compliance. \textbf{To design audit standards that effectively produce desired outcomes and support rigorous, outside assessments of compliance, we recommend that standards developers consider breaking requirements out into smaller units to ensure that the standard adequately accounts for crucial preliminary steps that other requirements rely on.}

\subsubsection{Engaging stakeholders beyond those who will be using the standard can help complement users’ perspectives and can uncover critical misunderstandings or disagreements over desired audit practices.}
\label{sec:disc-design-codesign}
As researchers and policymakers increasingly call for multistakeholder engagement in the design of audit standards~\cite{schor2024mind, aistandardshub} and RAI tooling more broadly, we argue that \textbf{it is crucial that co-design efforts engage groups beyond the standard’s intended users.} This can help ensure that users’ desires for flexibility and compatibility with practice do not undermine effectiveness (as discussed earlier in Section~\ref{sec:disc-compatibility}). 

Furthermore, \textbf{this broader engagement can help reveal how desired goals for audit practices may differ and conflict among individual members of standards-developing organizations (e.g., members of the ASB), users of standards (e.g., forensic laboratories), and various audit stakeholders (e.g., defense attorneys~\cite{jin2024beyond, abebe2022adversarial}, prosecutors, independent experts in forensic DNA profiling~\cite{krane2022using, butler2024dna})}. For instance, forensic laboratories may advocate for audit practices that make efficient use of laboratory resources,  molecular biology researchers may call for more experiments on large sample sizes and diverse samples, and defense attorneys may advocate for more thorough documentation practices. In the context of the U.S. criminal legal system, building on work by \citet{abebe2022adversarial} and \citet{jin2024beyond} that center the critical role of defense scrutiny of forensic software and call for adversarial audits~\cite{abebe2022adversarial}, we argue that it is especially crucial to engage defense attorneys' adversarial perspectives, as adversarialism is crucial to protecting defendants’ rights and the truth-seeking goals of the criminal legal system as a whole~\cite{abebe2022adversarial}. In other auditing contexts, engaging adversarial perspectives can not only help reveal standards developers' blindspots and make visible a broader range of concerns, but can also highlight where there is less consensus in audit scope, methods, and goals~\cite{geiger2024making, raji2022outsider, birhane2024ai}.

Frameworks that encourage fine-grained specification of desired outcomes can not only help standards bodies improve the specificity of their standards, towards creating effective audit standards, but can also help scaffold efforts to co-design standards with diverse stakeholders. Similar to how boundary objects for contested concepts like privacy~\cite{mulligan2016privacy} and fairness~\cite{mulligan2019thing} can support communication and collaboration across diverse groups of stakeholders, we suggest that using frameworks for auditing (e.g., \cite{abebe2022adversarial, ojewale2025towards, raji2020closing}) can facilitate better understanding of potential misunderstandings or disagreements over core concepts, practices, and desired outcomes. 

%% file: 7-conclusion.tex
\section{Conclusion}
This paper investigates how the design of an audit standard can undermine its effectiveness, complicate efforts to assess audit compliance with the standard, and undermine accountability through a case study of ASB 018, a standard for auditing probabilistic genotyping software. Through qualitative analysis of ASB 018 and five publicly available audit reports, we identify numerous gaps between the standard's desired outcomes and the auditing practices it enables. We identify a number of design features that enable these gaps, such as the standard's failure to define key concepts, requirements that treat audit components as separate considerations that can be assessed independently of one another, and vague language. Based on our findings, we recommend that designers of audit standards clearly articulate their goals, carefully examine how designing for compatibility with users' practices can undermine effectiveness, ensure sufficient specificity in the requirements language to ensure effectiveness and accountability, and co-design standards with stakeholders beyond the standard's intended users.

%% file: 8-appendix.tex
\section{Additional Background}
\subsection{The creation of the ASB}
\label{appendix-a}
The creation of the AAFS Standards Board (ASB) was in large part motivated by the 2009 National Academies report, which recommended, amongst other things, the creation of standards for forensic science to ``ensure desirable characteristics of services and techniques such as quality, reliability, efficiency, and consistency among practitioners'' and ``maintain autonomy from vested interest groups''~\cite{national2009strengthening}. As opposed to other organizations that responded to this call, the ASB was specifically created in response to interest in the forensic science community, in developing an ANSI-accredited standards-developing organization~\cite{ambrosius2025celebrating}. As the organization describes on their website, ``In July 2015, AAFS received a \$1.5 million grant from the Laura and John Arnold Foundation over a four-year period to become an accredited [standards-developing organization] and provide the standards to the public free of charge. Since then, the [ASB] has formed 15 discipline-specific Consensus Bodies covering everything from DNA and toxicology to forensic nursing and wildlife forensics''~\cite{ambrosius2025celebrating}.

\section{Additional Methods}
\label{sec:appendix-b}
\subsection{Our interpretations of ASB 018's line-level requirements that relate to internal validation}
\label{sec:appendix-b-interpretations}
Below, for each ASB 018 line-level requirement that we deem relevant for internal validation studies (quoted in bold text), we provide our interpretation of the requirement. We understood two of these requirements to be tautological, or always be met, regardless of the internal validation study's specific practices (R4.1.3-d, R4.1.3-e). For several others, we interpreted the requirement to be met simply by the existence of the audit and audit report. For both of these groups of requirements, we explain our interpretation under the quoted requirement. For all other requirements included in our study, we phrased our interpretation as a question that we used to assess each audit report for compliance with the requirement, which we document below. In Appendix~\ref{appendix-b-compliance}, we document sections of each audit report that we take to fulfill each of our interpretations of this third group of line-level requirements.

In constructing our interpretations, we drew on ASB 018's own definitions of terms when available. Throughout, we aimed to reduce ambiguity while retaining as much of ASB 018's language as possible.

\textbf{Notation:} When a single numbered ASB 018 requirement contains multiple line-level requirements, we add letters after the ASB 018 requirement number to label each line-level requirement. For example, Requirement 4.1.3 contains five individual ``shall'' statements, which we distinguish using letters (e.g., R4.1.3a-e). Several of the line-level requirements we include in our analysis are pulled from Annex A of ASB 018, which we denote using ``annex'' (e.g., R4.1.3-annex).

\begin{itemize}
    \item \textbf{(R4.1) ``The laboratory shall validate a probabilistic genotyping system prior to its use for casework samples in the laboratory.''}~\cite[p.~3]{asb018} 
    \begin{itemize}
        \item Does the report describe the laboratory as having at least some part in conducting or reviewing the internal validation study?
    \end{itemize}
    
    \item \textbf{(R4.1.1-a) ``Validations shall include both developmental and internal studies.''}~\cite[p.~3]{asb018} 
    \begin{itemize}
        \item \textit{Line-level requirement simply requires the existence of an internal validation study.}
    \end{itemize}
    
    \item \textbf{(R4.1.1-b) ``Developmental validation shall not replace internal validation.''}~\cite[p.~3]{asb018}
    \begin{itemize}
        \item \textit{Line-level requirement simply requires the existence of an internal validation study.}
    \end{itemize}

    \item \textbf{(R4.1.3-a) ``Internal validation studies shall address the following: accuracy, sensitivity, specificity, and precision.''}~\cite[p.~3]{asb018}   
    \begin{itemize}
        \item (i) Does any part of the report address the software's accuracy (i.e., the software's ability to produce the expected likelihood ratio as calculated manually or with an alternate software program or application)?
        \item (ii) Does any part of the report address the software's sensitivity (i.e., the software's ability to produce inclusionary likelihood ratios for known contributors)?
        \item (iii) Does any part of the report address the software's specificity (i.e., the software's ability to produce exclusionary likelihood ratios for known non-contributors)?
        \item (iv) Does any part of the report address the software's precision (i.e., the variation in likelihood ratios calculated from repeated software analyses of the same input data using the same set of conditions/parameters)?
    \end{itemize}
    
    \item \textbf{(R4.1.3-b) ``These studies shall include internally generated case-type profiles of known composition that represent (in terms of number of contributors, mixture ratios, and total DNA template quantities) the range of actual casework samples intended for analysis with the system at the laboratory.''}~\cite[p.~3]{asb018}
    \begin{itemize}
        \item Does the report include internally generated samples of known composition that span a range of actual casework samples (in terms of the number of contributors, mixture ratios, and total DNA template quantities)?
    \end{itemize}

     \item \textbf{(R4.1.3-c) ``Studies shall not be limited to pristine DNA samples but shall also include compromised DNA samples (e.g., low template, degraded, and inhibited samples).''}~\cite[p.~3]{asb018}
    \begin{itemize}
        \item Does the report include any DNA samples that the lab either describes as ``compromised'' or ``challenging'', or is low template, degraded, or inhibited?
    \end{itemize}

    \item \textbf{(R4.1.3-d) ``The internal validation shall not exceed the scope of conditions tested in the developmental validation.''}~\cite[p.~3]{asb018}
    \begin{itemize}
        \item \textit{Because ASB 018's broad definition of developmental validation could also encompass any internal validation, including the internal validation at hand, this requirement is always met.}
    \end{itemize}

    \item \textbf{(R4.1.3-e) ``Case type profiles that fall outside the range of conditions explored in the developmental validation shall require additional developmental validation studies.''}~\cite[p.~3]{asb018}
    \begin{itemize}
        \item \textit{Condition triggering this requirement is never met given ASB 018's broad definition of developmental validation (see notes for R4.1.3-d).}
    \end{itemize}

    \item \textbf{(R4.1.3-annex) ``The laboratory shall perform sufficient studies to address the variability inherent to the various aspects of DNA testing, data generation, analysis and interpretation of data and user input parameters.''}~\cite[p.~5]{asb018}
    \begin{itemize}
        \item Does any part of the report test the software using a set of at least two samples that captures at least one source of variability in DNA profile generation, DNA profile interpretation by users, or analysis?
    \end{itemize}

    \item \textbf{(R4.1.4-a) ``Internal validation studies shall include evaluating user input parameters that vary run to run.''}~\cite[p.~3]{asb018}
    \begin{itemize}
        \item Does the report include any experiment that changes a user input?
    \end{itemize}

    \item \textbf{(R4.1.4-b) ``The effects of artifacts (e.g., stutter) and parameters that relate to the statistical algorithm (e.g., run time parameters for the software system that can vary from system to system) shall also be evaluated. [...] [T]he specific parameters to be tested shall be determined by the laboratory.''}~\cite[p.~3]{asb018}
    \begin{itemize}
        \item (i) Does any part of the report assess the effect of an artifact on some aspect of system behavior?
        \item (ii) Does any part of the report assess the effect of a parameter that relates to the statistical algorithm on some aspect of system behavior? Are these parameters determined by the laboratory?
    \end{itemize}

    \item \textbf{(R4.1.5-a) ``Internal validation studies shall also include the evaluation of multiple propositions for case type samples to aid in the development of propositions.''}~\cite[p.~3]{asb018}
    \begin{itemize}
        \item Does any part of the report use multiple propositions?
    \end{itemize}
    
    \item \textbf{(R4.1.5-b) ``Such studies shall also consider the effect of overestimating and underestimating the number of contributors.''}~\cite[p.~3]{asb018}
    \begin{itemize}
        \item Does any part of the report consider the effect of overestimating and underestimating the number of contributors on some aspect of system behavior?
    \end{itemize}

    \item \textbf{(R4.1.6) ``For internal validation, the laboratory shall evaluate both the appropriate sample types (i.e., number of contributors, mixture ratios, and template quantities) and the number of samples within each type to demonstrate the potential limitations and reliability of the software. The laboratory shall base this evaluation on the intended application of the software.''}~\cite[p.~3]{asb018}
    \begin{itemize}
        \item Does any part of the report claim that the study's evaluated samples are appropriate for the intended application of the software?
    \end{itemize}

    \item \textbf{(R4.4-annex) ``Additional validation or a performance check shall be based on the list of documented changes provided by the developer that accompany each updated version of the software installed in the laboratory.''}~\cite[p.~5]{asb018}
    \begin{itemize}
        \item If the report indicates that the lab has adopted a previous version of the software, does any part of the report suggest the study is based on information about documented changes provided by the developer?
    \end{itemize}

    \item \textbf{(R4.5) ``All validation and performance check studies conducted by the laboratory shall be documented and retained by the laboratory.''}~\cite[p.~4]{asb018}
    \begin{itemize}
        \item \textit{Because all of the audits we analyze in this study have been documented in the audit reports we are analyzing, this requirement is met for all audits in our study.}
    \end{itemize}
\end{itemize}

\subsection{Compliance assessment}
In Table~\ref{appendix-table-compliance}, we document sections of each audit report that we take to fulfill each of our interpretations of ASB 018's line-level requirements (see Appendix~\ref{sec:appendix-b-interpretations} for our interpretations). 

\textbf{Additional note on our assessment of PBSO internal validation study of STRmix v2.6.2 against R4.4-annex:} While not explicitly describing differences between v2.6.2 and the previously validated v2.4, the PBSO v2.6.2 report does describe developer-provided documentation accompanying v2.6.2 as a guide for designing the PBSO validation of v2.6.2 and deliberately avoids re-validating certain software behaviors it deemed irrelevant to the study.

\label{appendix-b-compliance}
\begin{table*}[h]
\begin{tabular}{|  l  |l  |l|l  |l|l  |}
\hline
 \textbf{Line-level Req.} & \textbf{OCME v2.4 (2016)} &  \textbf{CBI v2.5 (2018)} &\textbf{PBSO v2.6.2 (2019)} &  \textbf{OCME v2.7 (2021)} &\textbf{MSP v2.9.1 (2024)}\\ 
 \hline\hline
 4.1 & Introduction (p. 2) &  Introduction (p. 2) &Introduction (p. 2) &  Introduction (p. 1) &Introduction (p. 3) \\  
 \hline
 4.1.3-a (i) & Exp. 2 (p. 3-5) &  Sec. A (p. 4-5) &Sec. A (p. 5-7) &  Exp. 2 (p. 4-6) &Sec. A (p. 5-7) \\
 \hline
 4.1.3-a (ii) & Exp. 4 (p. 7-17) &  Sec. D (p. 7-32) &Sec. D (p. 8-15) &  Exp. 4 (p. 14-25) &Sec. D (p. 13-29)  \\
 \hline
 4.1.3-a (iii) & Exp. 4 (p. 7-17) &  Sec. D (p. 7-32) &Sec. D (p. 8-15) &  Exp. 4 (p. 14-25) &Sec. D (p. 13-29)  \\
 \hline
 4.1.3-a (iv) & Exp. 6 (p. 19-22) &  Sec. M (p. 52-55) &Sec. M (p. 25-31) &  Exp. 6 (p. 28-30) &Sec. M (p. 64-66)\\
 \hline
 4.1.3-b & (p. 2, 40) &  (p. 8, 56)  &(p. 2, 31) &  (p. 2, 64) &(p. 3, 14)  \\
 \hline
 4.1.3-c & Exp. 14 (p. 38-40) &  Sec. L (p. 47-52) &Sec. L (p. 25) &  Exp. 13 (p. 54-63) &Sec. J (p. 48-58)  \\
 \hline
 4.1.3-annex & Exp. 9 (p. 24-27) &  Sec. M (p. 52-53) &Sec. F (p. 18-22) &  Exp. 6 (p. 28-30) &Sec. M (p. 64-66)  \\
 \hline
 4.1.4-a & Exp. 10 (p. 27-33) &  Sec. F (p. 35-38) &Sec. F (p. 18-22) &  Exp. 9 (p. 36-44) &Sec. F (p. 30-40) \\
 \hline
 4.1.4-b (i) & Exp. 11 (p. 33-35) &  Sec. G (p. 40) &Sec. G (p. 22) &  Exp. 10 (p. 44-48) &Sec. G (p. 41-44)  \\
 \hline
 4.1.4-b (ii) & Exp. 6 (p. 19-22) &  Sec. M (p. 53-55) &Appendix 3 (p. 36-45) &  Exp. 6 (p. 28-30) &Sec. M (p. 64-66) \\
 \hline
 4.1.5-a & Exp. 5 (p. 18-19) &  Sec. E (p. 32-34) &Sec. E (p. 16-17) &  Exp. 5 (p. 25-27) &Sec. E (p. 29-30)  \\
 \hline
 4.1.5-b & Exp. 10 (p. 27-33) &  Sec. F (p. 35-38) &Sec. F (p. 18-22) &  Exp. 9 (p. 36-44) &Sec. F (p. 30-40)  \\
 \hline
 4.1.6 & Conclusion (p. 40) &  Conclusion (p. 56) &Conclusion (p. 31) &  Conclusion (p. 64) &Conclusion (p. 96)  \\
 \hline
  4.4-annex & N/A &  N/A &Introduction (p. 2) &  Introduction (p. 1-2) &N/A\\
 \hline

\end{tabular}
\caption{For each internal validation report, we document the page numbers of the internal validation report that contain text that satisfies our interpretation of each line-level requirement. When multiple sections of a report could satisfy our interpretation of a given line-level requirement, we document the page numbers for one of the sections. All page numbers refer to the numbered pages of the internal validation report. The OCME v2.4 audit report was updated in 2019 to correct transcriptional errors and improve the clarity of tables, figures and end notes---the original audit was completed in 2016.}
\label{appendix-table-compliance}
\end{table*}

%% file: sample-base.bib
@String{Computing = "Computing" }

@String{Computer = "{IEEE} Computer" }

@String{Springer = "Springer-Verlag" }

@inproceedings{lam2024framework,
  title={A framework for assurance audits of algorithmic systems},
  author={Lam, Khoa and Lange, Benjamin and Blili-Hamelin, Borhane and Davidovic, Jovana and Brown, Shea and Hasan, Ali},
  booktitle={The 2024 ACM Conference on Fairness, Accountability, and Transparency},
  pages={1078--1092},
  year={2024}
}

@inproceedings{raji2020closing,
  title={Closing the AI accountability gap: Defining an end-to-end framework for internal algorithmic auditing},
  author={Raji, Inioluwa Deborah and Smart, Andrew and White, Rebecca N and Mitchell, Margaret and Gebru, Timnit and Hutchinson, Ben and Smith-Loud, Jamila and Theron, Daniel and Barnes, Parker},
  booktitle={Proceedings of the 2020 conference on fairness, accountability, and transparency},
  pages={33--44},
  year={2020}
}

@inproceedings{raji2022outsider,
  title={Outsider oversight: Designing a third party audit ecosystem for ai governance},
  author={Raji, Inioluwa Deborah and Xu, Peggy and Honigsberg, Colleen and Ho, Daniel},
  booktitle={Proceedings of the 2022 AAAI/ACM Conference on AI, Ethics, and Society},
  pages={557--571},
  year={2022}
}

@article{goodman2022algorithmic,
  title={Algorithmic Auditing: Chasing AI Accountability},
  author={Goodman, Ellen P and Trehu, Julia},
  journal={Santa Clara High Tech. LJ},
  volume={39},
  pages={289},
  year={2022},
  publisher={HeinOnline}
}

@inproceedings{groves2024auditing,
  title={Auditing work: Exploring the New York City algorithmic bias audit regime},
  author={Groves, Lara and Metcalf, Jacob and Kennedy, Alayna and Vecchione, Briana and Strait, Andrew},
  booktitle={The 2024 ACM Conference on Fairness, Accountability, and Transparency},
  pages={1107--1120},
  year={2024}
}

@inproceedings{wright2024null,
  title={Null Compliance: NYC Local Law 144 and the challenges of algorithm accountability},
  author={Wright, Lucas and Muenster, Roxana Mika and Vecchione, Briana and Qu, Tianyao and Cai, Pika and Smith, Alan and Comm 2450 Student Investigators and Metcalf, Jacob and Matias, J Nathan and others},
  booktitle={The 2024 ACM Conference on Fairness, Accountability, and Transparency},
  pages={1701--1713},
  year={2024}
}

@inproceedings{costanza2022audits,
  title={Who Audits the Auditors? Recommendations from a field scan of the algorithmic auditing ecosystem},
  author={Costanza-Chock, Sasha and Raji, Inioluwa Deborah and Buolamwini, Joy},
  booktitle={Proceedings of the 2022 ACM Conference on Fairness, Accountability, and Transparency},
  pages={1571--1583},
  year={2022}
}

@online{strmix2025use,
  author = {STRmix},
  title = {STRmix™ Now Being Used to Interpret Crime Scene Evidence in 91 U.S. Labs},
  url = {https://www.strmix.com/news/strmix-now-being-used-to-interpret-crime-scene-evidence-in-91-u-s-labs},
  urldate = {Sept 10, 2025},
  year = {Mar 11, 2025}
}

@online{ambrosius2025celebrating,
  author = {Teresa Ambrosius},
  title = {Celebrating Ten Years of the AAFS Standards Board},
  url = {https://www.aafs.org/article/celebrating-ten-years-aafs-standards-board},
  urldate = {Dec 5, 2025},
  year = {Nov 4, 2025}
}

@online{aistandardshub,
  author = {AI Standards Hub},
  title = {About the AI Standards Hub},
  url = {https://aistandardshub.org/the-ai-standards-hub/},
  urldate = {Jan 15, 2026},
  year = {2024}
}

@misc{butler2024dna,
	title = {{DNA} {Mixture} {Interpretation}: {A} {NIST} {Scientific} {Foundation} {Review}},
	url = {https://tsapps.nist.gov/publication/get_pdf.cfm?pub_id=959142},
	doi = {https://doi.org/10.6028/NIST.IR.8351},
	language = {en},
	publisher = {NIST Interagency/Internal Report (NISTIR), National Institute of Standards and Technology, Gaithersburg, MD},
	author = {Butler, John and Iyer, Hariharan and Press, Richard and Taylor, Melissa and Vallone, Peter and Willis, Sheila},
	month = dec,
	year = {2024},
}

@article{butler2024history,
  title={History of DNA Mixture Interpretation: Supplemental Document to DNA Mixture Interpretation: A NIST Scientific Foundation Review},
  author={Butler, John},
  year={2024},
  publisher={John Butler}
}

@online{ansi,
  author = {American National Standards Institute (ANSI)},
  title = {American National Standards (ANS) Introduction},
  url = {https://www.ansi.org/american-national-standards/ans-introduction/overview#introduction},
  urldate = {Jan 22, 2025}
}

@online{asb2025about,
  author = {American Academy of Forensic Sciences (AAFS)},
  title = {About ASB},
  url = {https://www.aafs.org/academy-standards-board/about-asb},
  urldate = {Sept 11, 2025}
}

@misc{2023anderson,
    author="{United States v. Anderson, 673 F. Supp. 3d 671 (21-CR-204)}",
    year={(M.D. Pa. 2023)}
}

@misc{2020lewis,
    author="{United States v. Lewis, 442 F. Supp. 3d 1122 (18-CR-194)}",
    year={(D. Minn. 2020)}
}

@misc{foley,
    author="{Commonwealth v. Foley, 38 A.3d 882, 888–90}",
    year={(Pa. Super. Ct. 2012)}
}

@misc{ortiz,
    author="{United States v. Ortiz, 736 F.Supp.3d 895 (21-CR-2503)}",
    year={(S.D. Cal. 2024)}
}

@misc{johnston,
    author="{United States v. Johnston, (23-CR-13)}",
    year={(E.D. N.Y. 2025)}
}

@misc{wexlaw_daubert,
    author="Legal Information Institute at Cornell University",
    title={Daubert Standard},
    year={2023},
    url={https://www.law.cornell.edu/wex/daubert_standard},
    urldate={January 22, 2025}
}

@article{roth2016machine,
  title={Machine testimony},
  author={Roth, Andrea},
  journal={Yale LJ},
  volume={126},
  pages={1972},
  year={2016},
  publisher={HeinOnline}
}

@article{obermeyer2019dissecting,
  title={Dissecting racial bias in an algorithm used to manage the health of populations},
  author={Obermeyer, Ziad and Powers, Brian and Vogeli, Christine and Mullainathan, Sendhil},
  journal={Science},
  volume={366},
  number={6464},
  pages={447--453},
  year={2019},
  publisher={American Association for the Advancement of Science}
}

@inproceedings{chouldechova2018case,
  title={A case study of algorithm-assisted decision making in child maltreatment hotline screening decisions},
  author={Chouldechova, Alexandra and Benavides-Prado, Diana and Fialko, Oleksandr and Vaithianathan, Rhema},
  booktitle={Conference on fairness, accountability and transparency},
  pages={134--148},
  year={2018},
  organization={PMLR}
}

@article{wu2021medical,
  title={How medical AI devices are evaluated: limitations and recommendations from an analysis of FDA approvals},
  author={Wu, Eric and Wu, Kevin and Daneshjou, Roxana and Ouyang, David and Ho, Daniel E and Zou, James},
  journal={Nature Medicine},
  volume={27},
  number={4},
  pages={582--584},
  year={2021},
  publisher={Nature Publishing Group US New York}
}

@techreport{asb018,
type = {Standard},
key = {ANSI/ASB Standard 018, 1st Ed.},
month = jul,
year = {2020},
title = {{Standard for Validation of Probabilistic Genotyping Systems}},
volume = {1},
address = {Colorado Springs, CO},
institution = {AAFS Standards Board},
url={https://www.aafs.org/sites/default/files/media/documents/018_Std_e1.pdf}
}

@techreport{asb018factsheet,
type = {Factsheet},
year = {2021},
title = {{Factsheet for ANSI/ASB Standard 018}},
institution = {American Academy of Forensic Sciences},
url = {https://www.aafs.org/sites/default/files/media/documents/ASB%20018%20DNA%20%28Revision%29.pdf}
}

@techreport{swgdam2015guidelines,
type = {Guidelines},
month = jun,
year = {2015},
title = {{Guidelines for the Validation of Probabilistic Genotyping Systems}},
institution = {Scientific Working Group on DNA Analysis Methods (SWGDAM)},
url={https://www.swgdam.org/_files/ugd/4344b0_22776006b67c4a32a5ffc04fe3b56515.pdf}
}

@techreport{enfsi2023guidelines,
type = {Guidelines},
month = oct,
year = {2023},
title = {{Guideline for Internal Validation / Verification of Various Aspects of the DNA Profiling Process}},
institution = {The European Network of Forensic Science Institutes (ENFSI)},
url={https://enfsi.eu/wp-content/uploads/2024/02/ENFSI-Validation-Guideline-04012024.pdf}
}

@techreport{ukfsr2024guidelines,
type = {Guidelines},
month = jul,
year = {2024},
title = {{Software Validation for DNA Mixture Interpretation}},
institution = {UK Forensic Science Regulator},
url={https://assets.publishing.service.gov.uk/media/5f607bbc8fa8f5106b23aa3a/G223_Mix_software_valid_Issue2_accessV3.pdf}
}

@techreport{fbi2025qas,
type = {Guidelines},
month = jul,
year = {2025},
title = {{Quality Assurance Standards for Forensic DNA Testing Laboratories}},
institution = {Federal Bureau of Investigation (FBI)},
url={https://www.swgdam.org/_files/ugd/4344b0_c2c9d0c7652f4977a57649ce500466aa.pdf}
}

@techreport{ocmeIV,
type = {Internal Validation Study Summary},
year = {2018},
title = {{Internal Validation of STRmix V2.4 for Fusion NYC OCME}},
institution = {NYC Office of the Medical Examiner (OCME)},
url = {https://www.nyc.gov/assets/ocme/downloads/pdf/STRmix-V2-4-Fusion-5C-Validation%20Summary.pdf}
}

@techreport{ocmeIV2.7,
type = {Internal Validation Study Summary},
year = {2021},
title = {{Internal Validation of STRmix v2.7 for Fusion 5C/3500xL Data}},
institution = {NYC Office of the Medical Examiner (OCME)},
url = {https://www.nyc.gov/assets/ocme/downloads/pdf/internal_validation_strmix_2_7.pdf}
}

@techreport{cbiIV,
type = {Internal Validation Study Summary},
year = {2018},
title = {{Internal Validation of STRmix V2.5 for the Colorado Bureau of Investigation (CBI) Forensic Laboratories (GlobalFiler, 3500xL CE)}},
institution = {Colorado Bureau of Investigation (OCME)},
url = {https://www.ascld.org/wp-content/uploads/formidable/43/2018-STRmix-Validation_FINAL-38656-1-Rev.pdf}
}

@techreport{pbsoIV,
type = {Internal Validation Study Summary},
year = {2019},
title = {{Palm Beach County Sheriff's Office Internal Validation of STRmix V2.6.2 (Powerplex Fusion6C, 3500xICE)}},
institution = {Palm Beach County Sheriff's Office (PBSO)},
url = {https://www.ascld.org/wp-content/uploads/formidable/43/PBSO-STRmix-v2.6.2-Internal-Validation-Powerplex-Fusion-6C-3500xl-CE.pdf}
}

@techreport{mspIV,
type = {Internal Validation Study Summary},
year = {2024},
title = {{Maryland State Police Forensic Sciences Division Internal Validation of STRmix V2.9.1)}},
institution = {Maryland State Police (MSP)},
url = {https://www.ascld.org/wp-content/uploads/formidable/43/FSD-Biology-STRmix-V2.9.1-Internal-Validation-Summary.pdf}
}

@inproceedings{radiya2023sociotechnical,
  title={A Sociotechnical Audit: Assessing Police Use of Facial rRecognition},
  author={Radiya-Dixit, Evani and Neff, Gina},
  booktitle={Proceedings of the 2023 ACM Conference on Fairness, Accountability, and Transparency},
  pages={1334--1346},
  year={2023}
}

@article{lam2023sociotechnical,
  title={Sociotechnical Audits: Broadening the Algorithm Auditing Lens to Investigate Targeted Advertising},
  author={Lam, Michelle S and Pandit, Ayush and Kalicki, Colin H and Gupta, Rachit and Sahoo, Poonam and Metaxa, Dana{\"e}},
  journal={Proceedings of the ACM on Human-Computer Interaction},
  volume={7},
  number={CSCW2},
  pages={1--37},
  year={2023},
  publisher={ACM New York, NY, USA}
}

@inproceedings{birhane2024ai,
  title={AI auditing: The broken bus on the road to AI accountability},
  author={Birhane, Abeba and Steed, Ryan and Ojewale, Victor and Vecchione, Briana and Raji, Inioluwa Deborah},
  booktitle={2024 IEEE Conference on Secure and Trustworthy Machine Learning (SaTML)},
  pages={612--643},
  year={2024},
  organization={IEEE}
}

@inproceedings{abebe2022adversarial,
  title={Adversarial scrutiny of evidentiary statistical software},
  author={Abebe, Rediet and Hardt, Moritz and Jin, Angela and Miller, John and Schmidt, Ludwig and Wexler, Rebecca},
  booktitle={Proceedings of the 2022 ACM Conference on Fairness, Accountability, and Transparency},
  pages={1733--1746},
  year={2022}
}

@incollection{krane2022using,
  title={Using Laboratory Validation to Identify and Establish Limits to the Reliability of Probabilistic Genotyping Systems},
  author={Krane, Dan E and Philpott, M Katherine},
  booktitle={Handbook of DNA Profiling},
  pages={297--319},
  year={2022},
  publisher={Springer}
}

@article{canellas2021defending,
  title={Defending IEEE software standards in federal criminal court},
  author={Canellas, Marc},
  journal={Computer},
  volume={54},
  number={6},
  pages={14--23},
  year={2021},
  publisher={IEEE}
}

@article{hoofnagle2016assessing,
  title={Assessing the Federal Trade Commission's Privacy Assessments},
  author={Hoofnagle, Chris Jay},
  journal={IEEE Security \& Privacy},
  volume={14},
  number={2},
  pages={58--64},
  year={2016},
  publisher={IEEE}
}

@article{coble2016dna,
  title={DNA Commission of the International Society for Forensic Genetics: Recommendations on the validation of software programs performing biostatistical calculations for forensic genetics applications},
  author={Coble, Michael D and Buckleton, John and Butler, John M and Egeland, T and Fimmers, R and Gill, P and Gusm{\~a}o, L and Guttman, B and Krawczak, Michael and Morling, N and others},
  journal={Forensic Science International: Genetics},
  volume={25},
  pages={191--197},
  year={2016},
  publisher={Elsevier}
}

@article{beyer1999contextual,
  title={Contextual design},
  author={Beyer, Hugh and Holtzblatt, Karen},
  journal={interactions},
  volume={6},
  number={1},
  pages={32--42},
  year={1999},
  publisher={ACM New York, NY, USA}
}

@misc{wyden2023algorithmic,
  author={Wyden, Sen. Ron},
  title={Algorithmic Accountability Act of 2023, S. 2892, 118th Congress},
  year={2023},
  note={\url{https://www.congress.gov/bill/118th-congress/senate-bill/2892}},
}

@inproceedings{ojewale2025towards,
author = {Ojewale, Victor and Steed, Ryan and Vecchione, Briana and Birhane, Abeba and Raji, Inioluwa Deborah},
title = {Towards AI Accountability Infrastructure: Gaps and Opportunities in AI Audit Tooling},
year = {2025},
isbn = {9798400713941},
publisher = {Association for Computing Machinery},
address = {New York, NY, USA},
url = {https://doi-org.libproxy.berkeley.edu/10.1145/3706598.3713301},
doi = {10.1145/3706598.3713301},
articleno = {815},
numpages = {29},
location = {
},
series = {CHI '25}
}

@inproceedings{buolamwini2018gender,
  title={Gender shades: Intersectional accuracy disparities in commercial gender classification},
  author={Buolamwini, Joy and Gebru, Timnit},
  booktitle={Conference on fairness, accountability and transparency},
  pages={77--91},
  year={2018},
  organization={PMLR}
}

@article{latonero2021human,
  title={Human rights impact assessments for AI: learning from Facebook’s failure in Myanmar},
  author={Latonero, Mark and Agarwal, Aaina},
  journal={Carr Center for Human Rights Policy Harvard Kennedy School, Harvard University},
  year={2021}
}

@inproceedings{gerchick2025auditing,
  title={Auditing the Audits: Lessons for Algorithmic Accountability from Local Law 144's Bias Audits},
  author={Gerchick, Marissa Kumar and Encarnaci{\'o}n, Ro and Tanigawa-Lau, Cole and Armstrong, Lena and Guti{\'e}rrez, Ana and Metaxa, Dana{\'e}},
  booktitle={Proceedings of the 2025 ACM Conference on Fairness, Accountability, and Transparency},
  pages={29--44},
  year={2025}
}

@article{selbst2021institutional,
  title={An institutional view of algorithmic impact assessments},
  author={Selbst, Andrew D},
  journal={Harv. JL \& Tech.},
  volume={35},
  pages={117},
  year={2021},
  publisher={HeinOnline}
}

@inproceedings{schor2024mind,
  title={Mind the gap: Designers and standards on algorithmic system transparency for users},
  author={Schor, Bianca GS and Norval, Chris and Charlesworth, Ellen and Singh, Jatinder},
  booktitle={Proceedings of the 2024 CHI Conference on Human Factors in Computing Systems},
  pages={1--16},
  year={2024}
}

@inproceedings{jackson2015standards,
  title={Standards and/as innovation: Protocols, creativity, and interactive systems development in ecology},
  author={Jackson, Steven J and Barbrow, Sarah},
  booktitle={Proceedings of the 33rd annual ACM conference on human factors in computing systems},
  pages={1769--1778},
  year={2015}
}

@inproceedings{berman2024scoping,
  title={A Scoping Study of Evaluation Practices for Responsible AI Tools: Steps Towards Effectiveness Evaluations},
  author={Berman, Glen and Goyal, Nitesh and Madaio, Michael},
  booktitle={Proceedings of the 2024 CHI Conference on Human Factors in Computing Systems},
  pages={1--24},
  year={2024}
}

@article{moss2015admissibility,
  title={The admissibility of TrueAllele: A computerized DNA interpretation system},
  author={Moss, Katherine L},
  journal={Wash. \& Lee L. Rev.},
  volume={72},
  pages={1033},
  year={2015},
  publisher={HeinOnline}
}

@inproceedings{lawrence2023bureaucratic,
  title={The bureaucratic challenge to AI governance: An empirical assessment of implementation at US federal agencies},
  author={Lawrence, Christie and Cui, Isaac and Ho, Daniel},
  booktitle={Proceedings of the 2023 AAAI/ACM Conference on AI, Ethics, and Society},
  pages={606--652},
  year={2023}
}

@book{national2009strengthening,
  title={Strengthening forensic science in the United States: a path forward},
  author={National Research Council and Division on Engineering and Physical Sciences and Committee on Applied and Theoretical Statistics and Global Affairs and Committee on Science and Law and Committee on Identifying the Needs of the Forensic Sciences Community},
  year={2009},
  publisher={National Academies Press}
}

@article{kelly2022exploring,
  title={Exploring likelihood ratios assigned for siblings of the true mixture contributor as an alternate contributor},
  author={Kelly, Hannah and Coble, Michael and Kruijver, Maarten and Wivell, Richard and Bright, Jo-Anne},
  journal={Journal of forensic sciences},
  volume={67},
  number={3},
  pages={1167--1175},
  year={2022},
  publisher={Wiley Online Library}
}

@inproceedings{madaio2020co,
  title={Co-designing checklists to understand organizational challenges and opportunities around fairness in AI},
  author={Madaio, Michael A and Stark, Luke and Wortman Vaughan, Jennifer and Wallach, Hanna},
  booktitle={Proceedings of the 2020 CHI conference on human factors in computing systems},
  pages={1--14},
  year={2020}
}

@inproceedings{metcalf2021algorithmic,
  title={Algorithmic impact assessments and accountability: The co-construction of impacts},
  author={Metcalf, Jacob and Moss, Emanuel and Watkins, Elizabeth Anne and Singh, Ranjit and Elish, Madeleine Clare},
  booktitle={Proceedings of the 2021 ACM conference on fairness, accountability, and transparency},
  pages={735--746},
  year={2021}
}

@inproceedings{deng2022exploring,
  title={Exploring how machine learning practitioners (try to) use fairness toolkits},
  author={Deng, Wesley Hanwen and Nagireddy, Manish and Lee, Michelle Seng Ah and Singh, Jatinder and Wu, Zhiwei Steven and Holstein, Kenneth and Zhu, Haiyi},
  booktitle={Proceedings of the 2022 ACM Conference on Fairness, Accountability, and Transparency},
  pages={473--484},
  year={2022}
}

@inproceedings{lee2021landscape,
  title={The landscape and gaps in open source fairness toolkits},
  author={Lee, Michelle Seng Ah and Singh, Jat},
  booktitle={Proceedings of the 2021 CHI conference on human factors in computing systems},
  pages={1--13},
  year={2021}
}

@inproceedings{mitchell2019model,
  title={Model cards for model reporting},
  author={Mitchell, Margaret and Wu, Simone and Zaldivar, Andrew and Barnes, Parker and Vasserman, Lucy and Hutchinson, Ben and Spitzer, Elena and Raji, Inioluwa Deborah and Gebru, Timnit},
  booktitle={Proceedings of the conference on fairness, accountability, and transparency},
  pages={220--229},
  year={2019}
}

@article{gebru2021datasheets,
  title={Datasheets for datasets},
  author={Gebru, Timnit and Morgenstern, Jamie and Vecchione, Briana and Vaughan, Jennifer Wortman and Wallach, Hanna and Iii, Hal Daum{\'e} and Crawford, Kate},
  journal={Communications of the ACM},
  volume={64},
  number={12},
  pages={86--92},
  year={2021},
  publisher={ACM New York, NY, USA}
}

@article{wong2023seeing,
  title={Seeing like a toolkit: How toolkits envision the work of AI ethics},
  author={Wong, Richmond Y and Madaio, Michael A and Merrill, Nick},
  journal={Proceedings of the ACM on Human-Computer Interaction},
  volume={7},
  number={CSCW1},
  pages={1--27},
  year={2023},
  publisher={ACM New York, NY, USA}
}

@inproceedings{holstein2019improving,
  title={Improving fairness in machine learning systems: What do industry practitioners need?},
  author={Holstein, Kenneth and Wortman Vaughan, Jennifer and Daum{\'e} III, Hal and Dudik, Miro and Wallach, Hanna},
  booktitle={Proceedings of the 2019 CHI conference on human factors in computing systems},
  pages={1--16},
  year={2019}
}

@article{rakova2021responsible,
  title={Where responsible AI meets reality: Practitioner perspectives on enablers for shifting organizational practices},
  author={Rakova, Bogdana and Yang, Jingying and Cramer, Henriette and Chowdhury, Rumman},
  journal={Proceedings of the ACM on Human-Computer Interaction},
  volume={5},
  number={CSCW1},
  pages={1--23},
  year={2021},
  publisher={ACM New York, NY, USA}
}

@article{mulligan2016privacy,
  title={Privacy is an essentially contested concept: a multi-dimensional analytic for mapping privacy},
  author={Mulligan, Deirdre K and Koopman, Colin and Doty, Nick},
  journal={Philosophical Transactions of the Royal Society A: Mathematical, Physical and Engineering Sciences},
  volume={374},
  number={2083},
  pages={20160118},
  year={2016},
  publisher={The Royal Society}
}

@article{mulligan2019thing,
  title={This thing called fairness: Disciplinary confusion realizing a value in technology},
  author={Mulligan, Deirdre K and Kroll, Joshua A and Kohli, Nitin and Wong, Richmond Y},
  journal={Proceedings of the ACM on Human-Computer Interaction},
  volume={3},
  number={CSCW},
  pages={1--36},
  year={2019},
  publisher={ACM New York, NY, USA}
}

@article{coble2019probabilistic,
  title={Probabilistic genotyping software: an overview},
  author={Coble, Michael D and Bright, Jo-Anne},
  journal={Forensic Science International: Genetics},
  volume={38},
  pages={219--224},
  year={2019},
  publisher={Elsevier}
}

@article{coble2015uncertainty,
  title={Uncertainty in the number of contributors in the proposed new CODIS set},
  author={Coble, Michael D and Bright, Jo-Anne and Buckleton, John S and Curran, James M},
  journal={Forensic Science International: Genetics},
  volume={19},
  pages={207--211},
  year={2015},
  publisher={Elsevier}
}

@article{paoletti2005empirical,
  title={Empirical analysis of the STR profiles resulting from conceptual mixtures},
  author={Paoletti, David R and Doom, Travis E and Krane, Carissa M and Raymer, Michael L and Krane, Dan E},
  journal={Journal of forensic sciences},
  volume={50},
  number={6},
  pages={JFS2004475--6},
  year={2005},
  publisher={American Society of Mechanical Engineers Digital Collection}
}

@inproceedings{jin2024beyond,
  title={(Beyond) Reasonable Doubt: Challenges that Public Defenders Face in Scrutinizing AI in Court},
  author={Jin, Angela and Salehi, Niloufar},
  booktitle={Proceedings of the 2024 CHI Conference on Human Factors in Computing Systems},
  pages={1--19},
  year={2024}
}

@inproceedings{matthews2020trusted,
  title={When trusted black boxes don't agree: Incentivizing iterative improvement and accountability in critical software systems},
  author={Matthews, Jeanna Neefe and Northup, Graham and Grasso, Isabella and Lorenz, Stephen and Babaeianjelodar, Marzieh and Bashaw, Hunter and Mondal, Sumona and Matthews, Abigail and Njie, Mariama and Goldthwaite, Jessica},
  booktitle={Proceedings of the AAAI/ACM Conference on AI, Ethics, and Society},
  pages={102--108},
  year={2020}
}

@inproceedings{matthews2019right,
  title={The right to confront your accusers: Opening the black box of forensic DNA software},
  author={Matthews, Jeanna and Babaeianjelodar, Marzieh and Lorenz, Stephen and Matthews, Abigail and Njie, Mariama and Adams, Nathaniel and Krane, Dan and Goldthwaite, Jessica and Hughes, Clinton},
  booktitle={Proceedings of the 2019 AAAI/ACM Conference on AI, Ethics, and Society},
  pages={321--327},
  year={2019}
}

@online{kinshipproblem,
  author = {Brooklyn Defender Services},
  title = {The Kinship Problem},
  url = {https://indefenseof.us/issues/kinship-problem},
  urldate = {Sept 10, 2025}
}

@article{sinha2022radically,
  title={Radically reimagining forensic evidence},
  author={Sinha, Maneka},
  journal={Ala. L. Rev.},
  volume={73},
  pages={879},
  year={2022},
  publisher={HeinOnline}
}

@misc{morrison2020vacuous,
  title={Vacuous standards--subversion of the OSAC standards-development process},
  author={Morrison, Geoffrey Stewart and Neumann, Cedric and Geoghegan, Patrick Henry},
  journal={Forensic Science International: Synergy},
  volume={2},
  pages={206--209},
  year={2020},
  publisher={Elsevier}
}

@article{geiger2024making,
  title={Making Algorithms Public: Reimagining Auditing From Matters of Fact to Matters of Concern},
  author={Geiger, Stuart R. and Tandon, Udayan and Gakhokidze, Anoolia and Song, Lian and Irani, Lilly},
  journal={International Journal of Communication},
  volume={18},
  pages={634-655},
  year={2024},
  url={https://ijoc.org/index.php/ijoc/article/download/20811/4455}
}
